\documentstyle[11pt,epsfig,fleqn]{article}
\title{
              A Phenomenological Description of
              \(\pi^{-}\Delta^{++}\)
              Photo- and Electroproduction
              in Nucleon Resonance Region}
\author{
M. Ripani\(^{a,1}\), V. Mokeev\(^{b}\), 
M. Anghinolfi\(^{a}\), M. Battaglieri\(^{a}\),\\
G. Fedotov\(^{d}\), E. Golovach\(^{a,b}\),
B.Ishkhanov\(^{b,d}\),  M. Osipenko\(^{d}\),\\
G. Ricco\(^{a,c}\), V. Sapunenko\(^{a,b}\),
M. Taiuti\(^{a}\) \\
\(^{}\)  \\
\(^{a}\)Istituto Nazionale di Fisica Nucleare, \\
Via Dodecanneso 33, I-16146 Genova (Italy) \\
\(^{}\)  \\
\(^{b}\)Nuclear  Physics  Institute,  Moscow  State  University, \\
Vorob'evy gory, 119899 Moscow, Russia \\
\(^{}\)  \\
\(^{c}\)Dipartimento di Fisica, Universit\`a di Genova, \\
Via Dodecanneso 33, I-16146, Genova, Italy \\
\(^{}\)  \\
\(^{d}\) Physical  Faculty  of  Moscow  State  University, \\
Vorob'evy gory, 119899 Moscow, Russia \\
}

\begin{document}
\maketitle

\begin{abstract}
The \(\pi^{-}\Delta^{++}\) production on the nucleon by real and virtual
photons is discussed as initial step in a simple model approach for
the two pion
photo- and electroproduction on the nucleon, with emphasis on nucleon resonance
excitation which is of interest for new facilities like TJNAF.
A calculation
for \(\pi^{-}\Delta^{++}\) channel in resonance excitation
region is presented and compared to existing experimental data along with
a discussion of physical effects that we find to be of relevance.
The calculation is proposed as a starting basis for the investigation of
\(N^{*}\) electromagnetic form factors using experimental data about
two pion production by real and virtual photons.
\end{abstract}

\footnotetext[1]{e-mail: \it ripani@ge.infn.it}

\section{Introduction.}
In  the  total photoabsorption cross section on the proton
above  400  MeV  the two pion production becomes possible, as it is
clearly shown by several data sets from bubble chamber real photon
experiments \cite{Bubble6768,LS71}, as well as from exclusive
electron scattering experiments \cite{Dam73,Wac78} and recent measurements
with real photon beams with the DAPHNE \cite{Bra95,Kru98} and
SAPHIR\cite{Kru98,Kle96} detectors.
From the data one can see that the cross section for  this
process shows a steep rise at photon energies above
threshold and then exhibits a smooth decrease, although still representing
a large fraction of the total hadron photoproduction at \(W > 1.8\) GeV. It  is
well  established \cite{Bubble6768,LS71,Dam73,Wac78,Kle96}
that a large contribution to this process
is  given  by  the intermediate formation of \(\Delta\)(1236), \(\rho\)(770)
and
their subsequent decay in the cascade processes
\begin{equation}
\gamma_{r,v} p \rightarrow \Delta^{++}\pi^{-} \rightarrow (p\pi^{+})\pi^{-}
\,\,\,\,\,
\end{equation}
\begin{equation}
\gamma_{r,v} p \rightarrow \rho^{0} p \rightarrow p(\pi^{+}\pi^{-})
\end{equation}
where indices \(r,v\) stand for real and virtual photons.

An important role in quasi two-body reactions (1),(2)
is  played by nucleon resonances, whose excitation is clearly
seen in the total photohadronic cross section \cite{Arm72} as well as in
exclusive channels like single pion photoproduction \cite{Wal69}.
Many  of such nucleon resonances  have  an
appreciable branching ratio in the \(\Delta \pi\) and \(\rho N\)
channels (see Table I)  and  for \(N^{*}\) heavier than 1.7 GeV
multipion decay channels become dominant. Therefore reactions (1),(2)
offer a promising opportunity to study the structure of high
lying resonances composed by light $u$ and $d$ quarks.
Moreover, quark model calculations
based on  the  three constituent quark picture \cite{Kon80,Cap94,Bij96}
predict more states
than those found in experimental searches\cite{PDG96}.
Some parameters for such "missing"
states  are  shown in Table II: many  of  them are expected to
have  a strong branching ratio in the two-pion final state and weak
or absent coupling to the single pion: this could be a reason for
them having escaped detection in experiments with
pion beams and in single pion photoproduction.
An additional prediction for these unobserved states is that
some of them could have sizeable electromagnetic couplings,
similar to other states observed in electromagnetic production.
Therefore, sizeable electromagnetic couplings and strong decays
through \(\Delta \pi\) or \(\rho N\) channels could make poorly
known states more visible as well as make some of
the "missing"  states appear in measurements of
photo- or electroproduction of the two-pion final states.
It is the aim of a wide experimental program
at new facilities like TJNAF \cite{Bur98,Rip98} to investigate
\(N^{*}\) structure and to perform a search for missing states
by measuring the multipion production exclusive channels produced
via electromagnetic interaction. In this paper we will focus
on the particular two-pion production channel that proceeds
through the \(\Delta^{++} \pi^{-}\) intermediate state.

An important feature in the study of resonances in reactions (1),(2) is
the strong contribution from non-resonant
processes \cite{LS71}, creating a "continuum" in which resonant states
are embedded.
For this reason, common  methods for resonance investigation
such as multipole amplitude analysis \cite{Ber75,Arn96}, which essentially
use resonance dominance, may be not very effective for reactions (1),(2).

Different approaches to describe double pion photoproduction have
been presented in a few papers\cite{LS71,Bar72,Mur96,Gom94,Och97}, 
where models based
on a variety of tree-level diagrams and a few baryon resonances
have been used to calculate the two pion production.
The limited number of resonances included however makes
them applicable only for W lower than about 1.6 GeV;
moreover non-resonant terms are not always
corrected for unitarity absorption effects: actually, a very important
problem in the investigation
of \(N^{*}\) in exclusive meson production by photons is the description of
non-resonant processes
at \(W > 1.6\) GeV, where many competitive channels open up
and coupled channel calculations\cite{Noz90,Sur95}
appear to be very difficult due to
restricted knowledge of hadron couplings as well as to computational
problems.

A phenomenological description of the interaction with open inelastic channels
in the initial and final states (ISI and FSI) has been proposed first
in \cite{LS71}.
In this approach particle absorption in the initial and final states accounts
for the complexity of interactions with inelastic channels; the absorptive
coefficients have been usually obtained\cite{LS71,Mur96,Got64} assuming a
diffractive character of the interaction of ingoing
and outgoing particles; this assumption could be justified at W values above
$N^{*}$ excitation region, while at $W < 2.0$ GeV other mechanisms like
especially s-channel processes should contribute in ISI and FSI significantly.
Moreover, the high value of the effective coupling coefficients,
close to the physical limit, adopted in  \cite{LS71,Mur96},
at relatively small $W \sim 1.6-1.7$ GeV looks
somewhat artificial and not physically justified.

The need to develop a method for \(N^{*}\)
electromagnetic form factor investigation from the experimental two-pion
photo- and electroproduction data led us to the
development of a phenomenological approach based on
minimal model ingredients. This was accomplished
parametrizing, under simple
assumptions, the main two-pion photo- and electroproduction
mechanisms and using experimental data to determine
the corresponding parameters.

This paper is our first step in  establishing  such
phenomenological approach: we focused our
attention on the particular quasi-two-body channel (1) and set up a
simple model including  all known resonant states
contributing to this reaction and non-resonant processes
starting from a minimal set of mechanisms proposed in \cite{LS71,Bar72}.
We used existing data for the pion
electromagnetic form factor to provide a description of non-resonant
terms for \(Q^{2} > 0\), as well as data about the strong form factor
in \(\pi N\Delta\)
vertex to take into account the particle size in hadronic
vertices. Unitarity effects manifesting in
the competition of many hadronic channels in non-resonant processes were
taken into account effectively, implementing initial and final state
interactions (ISI and FSI), in form of absorptive corrections;
a new feature of our approach is that we developed a specific
parametrisation of ISI and FSI mechanisms responsible for
channel coupling at W $\leq 2$ GeV, based on experimental
information about hadronic scattering amplitudes in this 
energy region. In this regard, we examined some aspects
of the \(N^{*}\) electromagnetic vertex dressing as related
to our implementation of higher order corrections in form
of ISI and FSI. We studied also the transition
to the higher energy region, W $>$ 2 GeV, where the
pure tree-level scheme corrected for finite
size (strong form factor) and unitarity effects (ISI-FSI)
fails in reproducing the experimental cross section:
we found that a Regge trajectory exchange, in the particular
form recently proposed\cite{Mur96,Gui97}, can be a valid 
continuation into the higher energy region.

Our model relates \(N^{*}\) electromagnetic
helicity  couplings \(A_{1/2}\), \(A_{3/2}\), \(C_{1/2}\) and
measured  cross sections for reaction (1) induced both by real and virtual
photons, therefore offering a way to attempt the measurement of
\(N^{*}\) contributions from a comparison to experimental data
or a fit.

\section{Helicity amplitudes and differential cross section.}
The helicity representation has been chosen to describe the amplitudes
for reaction (1).  The  differential
cross section for reaction (1) induced by  real  or  virtual
photons in the one-photon exchange approximation can be expressed
as \cite{Ama89}:
\begin{equation}
\frac{d\sigma}{d\Omega_{\pi}^{*}} =
\frac{1}{4K_{L}M_{N}}\left[ (4\pi\alpha)\frac{1-\varepsilon}{Q^{2}}
\frac{1}{2} L_{\mu\nu}W^{\mu\nu} \right]
\frac{1}{(2\pi)^{2}}\frac{p_{\pi}^{*}}{4W}
\end{equation}
\begin{eqnarray}
K_{L} = \frac{W^{2}-M_{N}^{2}}{2M_{N}},&      & \alpha = \frac{1}{137}
\nonumber
\end{eqnarray}
where \(W\) is CM total energy, \(p_{\pi}^{*}\) is the pion three momentum
modulus  in
the  CM  frame, \(K_{L}\) is the "equivalent" photon  energy, \(M_{N}\) is the
nucleon  mass, \(Q^{2}\) = - \(q^{2}\) where \(q^{\mu}\) is
the real-virtual photon
four-momentum, \(\varepsilon\) is the photon  polarization parameter,
 $L_{\mu\nu}$ is the leptonic tensor well-known from QED\cite{Ama89}.
The information about hadronic production is
contained  in the hadronic tensor \(W^{\mu\nu}\) which  is  a  bilinear
combination of hadronic currents:
\begin{equation}
W^{\mu\nu} = \frac{1}{2}\sum_{\lambda_{p}\lambda_{\Delta}}J_{\mu}^{*}J_{\nu}
\end{equation}
related to the helicity amplitudes \(\langle \lambda_{\Delta}
\left| T \right| \lambda_{\gamma}\lambda_{p} \rangle\) according to:
\begin{equation}
\varepsilon_{\mu}(\lambda_{\gamma})J^{\mu}(\lambda_{p}\lambda_{\Delta}) =
\langle \lambda_{\Delta} \left| T \right| \lambda_{\gamma}\lambda_{p}
\rangle
\end{equation}
where \(\lambda_{\Delta}\), \(\lambda_{p}\), \(\lambda_{\gamma}\) are
helicities of \(\Delta\), proton and photon, respectively,
$\varepsilon^{\mu}$ is the four -- vector
associated to the photon polarization state $\lambda_{\gamma}$
and all variables are evaluated in the reaction CM frame.

Being the \(\Delta\) an unstable particle, it is  necessary  to
fold (3) on its mass distribution as follows\cite{Pil79}:
\begin{equation}
\frac{d\sigma}{d\Omega_{\pi}^{*}} =
\int dM^{2}\frac{1}{\pi}
\frac{M_{\Delta}\Gamma_{\Delta}}{(M^{2}-M_{\Delta}^{2})^{2} +
M_{\Delta}^{2}\Gamma_{\Delta}^{2}}\frac{d\sigma}{d\Omega_{\pi}^{*}}(M^{2})
\end{equation}
where \(M^{2}\) is the squared running invariant mass of \(\Delta\),
while \(M_{\Delta}\),
\(\Gamma_{\Delta}\) are \(\Delta\)  mass and width, respectively.

\section{The tree-level diagrams.}
The mechanisms of reaction (1) were described by a minimal set of
Feynman tree-level diagrams presented in Fig. 1. These  diagrams
can  be  subdivided  into \(N^{*}\) contributions  (Fig.1a) and a
group of non-resonant processes, or Born terms  (Fig.1b-e). The
resonant part (Fig.1a) involves all relevant \(N^{*}\) and
\(\Delta^{*}\) excitations in  s-channel.
The non-resonant amplitudes correspond to the same set of mechanisms
considered
in \cite{LS71,Bar72}. New features of our approach in this respect are:
\begin{enumerate}
\item implementation of
electromagnetic vertex functions to describe the behaviour of
non-resonant amplitudes
at $Q^{2} > 0$ and
\item implementation of a strong $\pi N \Delta$ form factor,
relativistically invariant and based on the $N - N$ scattering
analysis, to take into account
the finite size of hadrons involved in non-resonant mechanisms. 
\end{enumerate}

\subsection{Non-resonant processes.}
The non-resonant processes are composed by the "seagull" or "contact"
term (Fig. 1b) (which also naturally arises when considering the
\(\pi N\Delta\) vertex as described by an effective meson-baryon
Lagrangian and then introducing the interaction with the electromagnetic
field by the minimal coupling prescription),
the  t-channel
pion-in-flight diagram (Fig.1c), the  u-channel delta-in-flight
diagram  (Fig.1e)  and by the  s-pole nucleon term  
(Fig.1d)\cite{LS71,Bar72}.
To treat photon, pion and delta off-shell we introduced
form factors in the corresponding vertices. Therefore Born terms
(Fig.1b-e) are functions of \(Q^{2}\) and
Mandelstam  t variables.
For the vertex function evaluation we used a compilation of
experimental data about electromagnetic and strong form
factors \cite{Ama89,Mac79,Beb78,Glo98};
these data provide a reliable vertex function evaluation in the relevant
region of \(Q^{2}\) and  t (\(Q^{2}< \sim 2\) \(GeV^{2}\),
\(t< \sim 2\) \(GeV^{2}\)).

In momentum space, helicity amplitudes corresponding
to the contact term are given by
\begin{equation}
f_{\lambda_{\Delta}\lambda_{\gamma}\lambda_{p}}^{c}(W,\theta) =
g_{c}(Q^{2},t)\overline{u}_{\mu}(p_{2},\lambda_{\Delta})
u(p_{1},\lambda_{p})\varepsilon_{\mu}(q,\lambda_{\gamma})
\end{equation}
where W is the usual invariant CM energy, $\theta$ is the pion production
angle in the hadronic CMS, \(p_{1}\) and \(p_{2}\) are the target nucleon and
\(\Delta\)-particle four momenta, \(q\) is the photon four momentum and
\(u_{\mu}\),
\(\varepsilon_{\mu}\) and \(u\) are
Rarita-Schwinger spinor-tensor for \(\Delta\)-particle with
\(\lambda_{\Delta}\)
helicity,
polarisation vector for photon with \(\lambda_{\gamma}\) helicity and spinor
for target nucleon with \(\lambda_{p}\) helicity, respectively;
\(g_{c}(Q^{2},t)\) is an effective "seagull" term vertex function.

The pion-in-flight contribution in momentum  representation
reads:
\begin{eqnarray}
f_{\lambda_{\Delta}\lambda_{\gamma}\lambda_{p}}^{pif}(W,\theta)
= \nonumber \\
g_{\pi}(Q^{2},t)\frac{(2p_{\pi}^{\mu}-
q^{\mu})\varepsilon_{\mu}(q,\lambda_{\gamma})}
{t-m_{\pi}^{2}}\overline{u}_{\nu}(p_{2},\lambda_{\Delta})u(p_{1},\lambda_{p})
(q^{\nu}-p_{\pi}^{\nu})
\end{eqnarray}
where \(p_{\pi}\) is the pion momentum, \(m_{\pi}\) is the pion mass,
\(g_{\pi}\) is the product
of strong and electromagnetic vertex functions:
\begin{equation}
g_{\pi}(Q^{2},t) = G_{\pi ,em}(Q^{2})G_{\pi N\Delta}(t)
\end{equation}
The electromagnetic vertex function in this case was described by the
well-known pole fit of pion form factor \cite{Ama89}
\begin{equation}
G_{\pi ,em}(Q^{2}) = \frac{1}{\left( 1+\frac{Q^{2}(GeV^{2})}{\Lambda_{\pi}^{2}}
\right) } \frac{1}{G_{\pi N \Delta}(t_{min})}
\end{equation}
 where $t_{min}$ corresponds to pion production in hadronic CMS at
zero degree angle; the factor $\frac{1}{G_{\pi N \Delta}(t_{min})}$
reflects the way of $G_{\pi e m}(Q^{2})$ form factor extraction from
single pion electroproduction data presented in \cite{Beb78}.  
Analysis in \cite{Beb78} yielded \(\Lambda_{\pi}^{2} = 0.462\) \(GeV^{2}\)
and we used this value in calculations. However considering uncertainties
in \cite{Beb78} as well as other data about pion electromagnetic
form factor \cite{Bra77}, we could allow for a variation of
$\Lambda_{\pi}^{2}$ cut-off
parameter within $0.4 - 0.5$ GeV$^{2}$. Concerning the $\pi N \Delta$
t-dependence, we introduce as vertex function a
strong form factor successfully applied in \(NN \rightarrow N\Delta\)
relativistic transition potentials\cite{Mac79,Mac87}
\begin{equation}
G_{\pi N\Delta}(t) = g_{0}\frac{\Lambda^{2}-m_{\pi}^{2}}{\Lambda^{2}-t}
\end{equation}
The interaction constant \(g_{0}\) and cut-off parameter \(\Lambda\) are:
\(g_{0}=2.1/m_{\pi}\) and \(0.6 < \Lambda < 0.9\) GeV\cite{Mac79,Glo98}.
We found that best data reproduction corresponds to $\Lambda = 0.75$ GeV.
Gauge invariance implies equality of the "seagull" term coupling 
\(g_{c}(Q^{2},t)\) 
and the pion-in-flight term coupling  \(g_{\pi}(Q^{2},t)\);
we stress that, instead of applying specific electromagnetic 
and strong vertex functions to the contact term (which would
imply to introduce some free parameters to be fitted), we
rather fixed the $\em product$ of them \(g_{c}(Q^{2},t)\)
requiring gauge invariance.

The   s-channel nucleon contribution (diagram of  Fig.1d)  is
given by
\begin{equation}
f_{\lambda_{\Delta}\lambda_{\gamma}\lambda_{p}}^{N}(W,\theta) =
g_{N}(Q^{2})g_{0}\frac{(2p_{1}^{\mu}+q^{\mu})\varepsilon_{\mu}(q,\lambda_{\gamma
})}
{s-m_{N}^{2}}\overline{u}_{\nu}(p_{2},\lambda_{\Delta})u(p_{1},\lambda_{p})
p_{\pi}^{\nu}
\end{equation}
where  s is Mandelstam invariant \(s=(q+p_{1})^{2}\), \(m_{N}\)
is the nucleon mass, \(g_{N}\) is
the electromagnetic vertex function (we
put the strong vertex  function for  s-channel equal to unity),
described
by the well-known dipole fit\cite{Ama89}:
\begin{equation}
g_{N}(Q^{2}) = \frac{1}{\left( 1+\frac{Q^{2}(GeV^{2})}{0.71} \right)^{2} }
\end{equation}

The  last contribution to the Born terms that we
considered was the \(\Delta\)-in-flight (diagram of Fig.1e). The
expression for such process is:
\begin{equation}
f_{\lambda_{\Delta}\lambda_{\gamma}\lambda_{p}}^{\Delta}(W,\theta) =
2g_{\Delta}(Q^{2},t)\frac{(2p_{2}^{\mu}-
q^{\mu})\varepsilon_{\mu}(q,\lambda_{\gamma})}
{u-m_{\Delta}^{2}}\overline{u}_{\nu}(p_{2},\lambda_{\Delta})p_{\pi}^{\nu}
u(p_{1},\lambda_{p})
\end{equation}
where u is Mandelstam variable corresponding to the crossed
invariant momentum transfer \(u=(p_{1}-p_{\pi})^2\);
factor 2 accounts for the double
electric charge of the \(\Delta^{++}\) particle, while again
\(g_{\Delta}(Q^{2},t)\)
is the product of
electromagnetic and strong vertex functions. Due to lack of
information about \(\Delta\)'s elastic electromagnetic form factors we
followed two ways to evaluate \(g_{\Delta}(Q^{2},t)\) vertex function:
first, it was derived from gauge invariance of
the total  Born amplitude:
\begin{equation}
g_{\Delta}(Q^{2},t) = \frac{g_{\pi}(Q^{2},t)+g_{N}(Q^{2})}{2} ;
\end{equation}
second, we used function (11) for $\pi N \Delta$ vertex, 
substituting
mass \(m_{\pi}\) by \(m_{\Delta}\). In the latter choice we found the
\(\Delta\)-in-flight contribution to be negligible,
because according to \cite{Mac79,Mac87}
\(\Lambda\) cut-off parameter does not exceed 1.3 GeV. Of course in the
second evaluation of \(\Delta\)-in-flight term gauge invariance
of the total Born amplitude is lost
but, considering that our description of non-resonant terms
aims to be phenomenological, we can assume
that gauge invariance could be restored by other mechanisms
not contributing significantly to the cross section.
Moreover, a negligible \(\Delta\)-in-flight contribution seems
to be supported by the comparison with the experimental data.

\subsection{The absorptive corrections.}
It is well known \cite{LS71} that the tree-level calculation is not
able to reproduce the data extracted from experimental
analysis of \(\pi^{-}\Delta^{++}\) production.
One reason is that the lowest order mechanisms for reaction (1)
do not incorporate unitarity and hence
interaction with other open channels in initial and final states must be
taken into account. Direct description of
the unitarity effects in the initial and final states
would require a simultaneous
treatment of all relevant hadronic channels in a coupled  channel
calculation using hadronic elastic and inelastic scattering
amplitudes data. The present status of the hadronic amplitudes knowledge
as well as computational capabilities restrict coupled
channel calculations to W lower than 1.6 GeV\cite{Noz90,Sur95}, while
investigation of the cross section for process (1) above
1.6 GeV is very important for \(N^{*}\) physics.

The description of open channels competition in non-resonance
processes remains a challenging open problem of \(N^{*}\) structure
investigation in exclusive meson photo- and electroproduction.
The modern status of strong interaction theory does not allow to
evaluate hadronic amplitudes and couplings from fundamental principles
(i.e. QCD Lagrangian). Therefore we chose to adopt an effective treatment
of unitarity effects in initial and final state;
the initial and final state interactions responsible for the modification
of the amplitude for reaction (1) are schematically depicted in
Fig. 2.
In the spirit of Vector Meson Dominance (VDM)\cite{LS71}, the
ISI was assumed to proceed through a transition between photon
and vector meson $\rho $.
The black blobs describe all transition processes, between $\rho p$ initial
and $\rho^{'} p^{'}$ intermediate states as well as between ${\pi^{-'}} 
{\Delta^{++'}}$ intermediate and $\pi^{-} \Delta^{++}$ final states.

We evaluated ISI and FSI effects in
\(\gamma p\rightarrow\pi^{-}\Delta^{++}\) reaction in the frame of
a simple phenomenological
recipe described in \cite{Got64}, where ingoing and outgoing particle
interaction is described by penetration
factors in the initial and final states \(f_{ISI}^{j}\), \(f_{FSI}^{j}\),
which depend on the reaction total angular momentum $j$; 
they 
determine the amplitude for $\rho p$ initial state to be transformed
into intermediate $\rho^{'} p^{'}$ state and the amplitude
for $\pi^{-'} \Delta^{++'}$ intermediate state to be transformed
into final $\pi^{-} \Delta^{++}$ state. 
According to \cite{Got64,Jac59}, the total Born amplitude
\(f_{\lambda_{\Delta}\lambda_{\gamma}\lambda_{p}}(\theta ,\varphi)\)
was numerically decomposed in partial waves with
total angular momentum  $j$:
\begin{eqnarray}
f_{\lambda_{\Delta}\lambda_{\gamma}\lambda_{p}}(\theta ,\varphi) &=&
\sum_{j}f_{\lambda\mu}^{j}d_{\lambda\mu}^{j}(\theta)e^{i(\lambda -\mu)\varphi}
\nonumber \\
f_{\lambda\mu}^{j} &=& \int d\Omega\frac{2j+1}{4\pi}
f_{\lambda_{\Delta}\lambda_{\gamma}\lambda_{p}}^{*}(\theta ,\varphi)
d_{\lambda\mu}^{j}(\theta)e^{i(\lambda -\mu)\varphi}
\\
\lambda = -\lambda_{\Delta}, &   & \mu = \lambda_{\gamma}-\lambda_{p}
\nonumber
\end{eqnarray}
The ISI \& FSI modify the $f^{j}_{\lambda \mu}$ decomposition
coefficients as:
\begin{eqnarray}
f^{j corr}_{\lambda \mu} = f^{j}_{ISI}f^{j}_{\lambda \mu}f^{j}_{FSI} \\
\lambda = \lambda_{\gamma} - \lambda_{p}, \mu = - \lambda_{\Delta}
\nonumber
\end{eqnarray}
where $f^{j corr}_{\lambda \mu}$ represent decomposition coefficients
for the Born term amplitude modified by ISI \& FSI.
$f^{j}_{ISI}$ and $f^{j}_{FSI}$ factors, 
under the assumptions of ref. \cite{Got64},
can be related
with $S^{j}$-matrix elements of
\(\pi^{-}\Delta^{++}\) and
\(\rho p\) elastic scattering amplitudes at total angular
momentum $j$ as:
\begin{eqnarray}
f_{ISI}^{j} &=& \langle\lambda_{\rho}\lambda_{p} |S^{j}|
\lambda_{\rho}\lambda_{p} \rangle^{1/2}
\nonumber \\
f_{FSI}^{j} &=& \langle \pi \lambda_{\Delta}|S^{j}|
\lambda_{\Delta} \pi \rangle^{1/2}
\end{eqnarray}
where \(\lambda_{\rho}\), \(\lambda_{p}\), \(\lambda_{\Delta}\)
are helicities for \(\rho\)-meson (equal to photon helicity in the VDM 
picture), for proton
and \(\Delta\)-isobar,
respectively;  S and T matrix elements are related as:
\begin{equation}
S = 1 + 2iT
\end{equation}

The penetration coefficients $f^{j}_{ISI}$ and $f^{j}_{FSI}$ are
unambiguously determined by $\pi^{-} \Delta^{++}$ and $\rho p$
elastic scattering amplitudes, therefore these amplitudes effectively
describe all mechanisms (resonant and non-resonant) responsible for
transitions between the initial $\rho p$ and intermediate $\rho^{'} p^{'}$
states as well as between the intermediate $\pi^{-'} \Delta^{++'}$ and
final $\pi^{-} \Delta^{++}$ states (blobs in Fig. 2).

As mentioned in the introduction, a new feature of our approach
is that, instead of using a diffractive ansatz like in the previous
literature\cite{Got64,LS71,Mur96}, better suited for higher energies,
T-matrix elements for \(\pi^{-}\Delta^{++}\) and \(\rho p\) 
elastic scattering
were more appropriately evaluated using data about hadronic scattering
in the resonance region, as briefly 
described in the following paragraphs.

T-matrix elements for
\(\pi^{-}\Delta^{++}\) and
\(\rho p\) elastic scattering were  evaluated in an isobar
model \cite{BatMSU,Bat98} assuming them to be a
superposition of relevant \(N^{*}\) contributions and a smooth background as
depicted in Fig. 3.
The resonant part of
the corresponding amplitudes was evaluated 
in Breit -- Wigner ansatz described in details in sect. 3.3.
To obtain $T_{res}$ amplitude normalization, we considered
a schematical situation, when complexity of all transitions
$\rho p \rightarrow \rho^{'} p^{'}$ and $\pi^{-'} \Delta^{++'}
\rightarrow \pi^{-} \Delta^{++}$ processes is limited to a single
$N^{*}$ excitation with elastic scattering as the only open channel;
for this case, ingoing and outgoing particle absorption should be
absent at resonant point ($W = M_{N^{*}}$) and the moduli of 
$f^{j}_{ISI}$ and $f^{j}_{FSI}$  coefficients should be equal to unity;
from unitarity conditions and from the standard resonance
phase behavior,
we determined $T^{j}_{res}$ amplitude normalization as
$\langle f |T^{j}_{res}| i \rangle =i$ at $W=M_{N^{*}}$, therefore
\begin{equation}
f^{j}_{ISI}, f^{j}_{FSI} = 1 + 2i \langle f |T^{j}_{res}| i \rangle = -1
\end{equation}
This way we obtained the following
relations between elastic $\pi \Delta$ and $\rho p$ scattering resonant
amplitudes
$\langle \pi \lambda_{\Delta}|T^{j}_{res}|\pi \lambda_{\Delta} \rangle$,
$\langle \lambda_{\rho} \lambda_{p}|T^{j}_{res}|
\lambda_{\rho}\lambda_{p} \rangle$
and $N^{*}$ decay helicity amplitudes
$\langle \pi \lambda_{\Delta}|T| N^{*} \rangle$,
$\langle \lambda_{\rho} \lambda_{p}|T| N^{*} \rangle$:
\begin{eqnarray}
\langle \pi \lambda_{\Delta} (\lambda_{\rho} \lambda_{p})  |T^{j}_{res}|
\pi \lambda_{\Delta} (\lambda_{\rho} \lambda_{p}) \rangle = 
\nonumber \\
\sum_{N^{*}} \left [ \frac{a^{j2}_{\lambda_{\Delta} (\lambda_{\rho} \lambda_{p})}}
{M^{2}_{N^{*}} - W^{2} - i \Gamma_{M_{N^{*}}}(W)M_{N^{*}}} \right ]
\left [ \frac{P^{c}_{\pi(\rho)}}{8\pi (2j + 1) W} \right]
\end{eqnarray}
where $M_{N^{*}}$, $\Gamma_{N^{*}}(W)$ are masses and W -- dependent
$N^{*}$ decay widths, $P^{c}_{\pi}$, $P^{c}_{\rho}$ are three -- momenta
moduli of pion and $\rho$.
Summation in (21) is performed over all $N^{*}$ contributing to the
partial wave
of total angular momentum $j$ and presented in table III;
$a^{j}_{\lambda_{\Delta}}$, $a^{j}_{\lambda_{\rho}\lambda_{p}}$ are
decomposition coefficients of 
$\langle \pi \lambda_{\Delta}|T| N^{*} \rangle$ and
$\langle \lambda_{\rho} \lambda_{p}|T| N^{*} \rangle$ $N^{*}$
decay amplitudes through the states of total angular momentum j and
related with $N^{*}$ partial decay widths in helicity representation
$\Gamma_{\lambda_{\Delta}}$ and  $\Gamma_{\lambda_{\rho} \lambda_{p}}$ 
according to (31).
Partial decay widths were taken from analysis \cite{Man92}
and transformed from LS to helicity representation.

Our approach should be somewhat modified if dressed $\gamma p N^{*}$ 
electromagnetic vertices are used for the resonant part of reaction (1)
amplitude (sect 3.3). 
Indeed, in the complexity of transition mechanisms in ISI \& FSI
shown in Fig. 2 by black blobs, also the sequence of
processes reported in Fig. 4 is effectively taken into account;
that clearly represents a dressing mechanism of the $\gamma p N^{*}$
electromagnetic vertex. But for
instance, $N^{*}$ electromagnetic helicity amplitudes $A_{1/2}$, $A_{3/2}$
extracted from pion photoproduction data analysis\cite{PDG96} are
usually interpreted as corresponding to dressed $\gamma p N^{*}$ verticies.
Therefore, a double counting problem can arise in this case, 
as dressing effects of the electromagnetic $N^{*}$ vertex can be
sizeable, as discussed in \cite{Koc84,Lee98}.
To avoid double
counting, one possibility is to exclude
$\rho p \rightarrow N^{*} \rightarrow \rho^{'} p^{'}$ and
$\pi^{-'} \Delta^{++'} \rightarrow N^{*} \rightarrow \pi^{-} \Delta^{++}$ 
mechanisms from ISI \& FSI.  
To work out a phenomenological prescription for 
such exclusion we considered the above mentioned schematical situation,
where the complexity of 
$\rho p \rightarrow \rho^{'} p^{'}$ and
$\pi^{-'} \Delta^{++'} \rightarrow \pi^{-} \Delta^{++}$ transition 
mechanisms is restricted to single $N^{*}$ excitation with only
elastic scattering as open channel. Unitarity for Breit -- Wigner
formula can be expressed as
\begin{equation}
\Gamma_{tot} = \sum_{i}{\Gamma_{i}},
\end{equation}
where $\Gamma_{tot}$, $\Gamma_{i}$ are total and partial $N^{*}$
decay widths.
For single $N^{*}$ with only elastic open channel
\begin{equation}
\Gamma_{tot} = {\Gamma_{el}},
\end{equation}
where $\Gamma_{tot}$, $\Gamma_{el}$ are total and elastic $N^{*}$
decay width; our assumption insures that all transition mechanisms in
$\rho p \rightarrow \rho^{'} p^{'}$,
$\pi^{-'} \Delta^{++'} \rightarrow \pi^{-} \Delta^{++}$
are represented by 
$\rho p \rightarrow N^{*} \rightarrow \rho^{'} p^{'}$ and
$\pi^{-'} \Delta^{++'} \rightarrow N^{*} \rightarrow \pi^{-} \Delta^{++}$
processes; therefore exclusion of $N^{*}$ excitation from ISI \& FSI
should lead to complete absorption of ingoing and outgoing particle or to
zero values for $f^{j}_{ISI}$, $f^{j}_{FSI}$ penetration factors.
As it follows from (19) the substitution
\begin{equation}
T^{j}_{res} \rightarrow \frac{1}{2} T^{j}_{res}
\end{equation}
would correspond to vanishing $f^{j}_{ISI}$, $f^{j}_{FSI}$
coefficients and, therefore, could be considered as an empirical 
prescription for excluding
$\rho p \rightarrow N^{*} \rightarrow \rho^{'} p^{'}$ and
$\pi^{-'} \Delta^{++'} \rightarrow N^{*} \rightarrow \pi^{-} \Delta^{++}$
mechanisms from ISI \& FSI treatment.

The complete absorption of Born term amplitudes takes place only for
such schematical situation (single $N^{*}$ excitation with only elastic
open channel) on which we based our exclusion prescription. 
In actual situation prescription (24) gives only partial absorption of 
ingoing and outgoing particles, since $N^{*}$ contributing to
$\pi^{-} \Delta^{++}$ and $\rho p$ elastic scattering have many open
decay channels, while relations (21) -- (24) give zero $f^{j}_{ISI}$,
$f^{j}_{FSI}$ coefficients only in our schematic assumption.
Inelasticities ($\Gamma_{tot} \neq \Gamma_{el}$) provide instead non-zero 
$f^{j}_{ISI}$, $f^{j}_{FSI}$ coefficients even at resonant point;
$N^{*}$ off-shell excitations as well as non-resonant processes also give
non-zero values of $f^{j}_{ISI}$ and $f^{j}_{FSI}$ coefficients.
We assumed that this partial absorption of Born terms 
can represent ISI \& FSI corrections, being the contribution from 
$\rho p \rightarrow N^{*} \rightarrow \rho^{'} p^{'}$ and
$\pi^{-'} \Delta^{++'} \rightarrow N^{*} \rightarrow \pi^{-} \Delta^{++}$
mechanisms excluded.

Of course our exclusion procedure (substitution (24)) 
should be applied only if for the
resonant part of reaction (1) dressed electromagnetic $\gamma p N^{*}$ 
vertices are used.
Actually, $\gamma p N^{*}$ vertices calculated in quark models 
are generally assumed to be ``bare'', i.e.
free from higher order corrections (see for
instance \cite{Cap94,Gia98}): in this case our calculation
should not be affected by any double counting and therefore no exclusion
procedure is necessary. The information on
$N^{*}$ electromagnetic form factors from the measured cross section
could be extracted both with and without application of exclusion procedure:
in the first case we would obtain information about dressed $\gamma p N^{*}$
vertices; in the second about $\gamma p N^{*}$ vertices without
contribution from higher order corrections depicted in Fig. 4.
Therefore our approach provides the flexibility of choosing
one procedure or the other.
 
Evaluation of non-resonant part of
$\pi^{-} \Delta^{++}$ and $\rho p$
scattering amplitude is described in more detail in 
\cite{BatMSU,Bat98},
but we want to report here the main features of our evaluation.
We used data\cite{Man92}  on partial-wave
$\pi N$ elastic
cross-sections with definite total angular momentum and isospin;
the amplitudes for these partial waves were described by a superposition of all
relevant $N^{*}$ (table 3) and a particular background for each
orbital momentum L and total spin S. The resonant part of $\pi N$
elastic scattering amplitudes was treated in the same manner as $\pi^{-}
\Delta^{++}$ and $\rho p$ elastic scattering processes; data\cite{Man92}
were used for \(N^{*}N\pi \) couplings. Background for each LS wave
was parametrized as a function of W by a simple linear dependence
\begin{equation}
T_{backgr LS}^{j} = A_{LS}W + B_{LS}
\end{equation}
Parameters $A_{LS}$ and $B_{LS}$ were determined from our fit of
$\pi N$ elastic scattering data\cite{BatMSU,Bat98}. To calculate background
amplitudes
for $\pi^{-} \Delta^{++}$ and $\rho p$ elastic scattering
starting from $\pi N$
elastic scattering background we used SU(3) flavour symmetry
relations\cite{BatMSU,Bat98}.
The fit results for $\pi N$ partial elastic scattering cross -- section
are shown in
Fig. 5. Dotted curves represent resonant contributions, dashed lines are
background contributions, while solid lines correspond to complete
amplitudes. As follows from Fig. 5, in $N^{*}$ excitation region nucleon
resonances provide the main contribution in $\pi N$ elastic amplitudes,
therefore also in penetration coefficients (17).
This contribution was completely neglected in previous evaluations
\cite{LS71,Mur96}.
Hence, implementation of s -- channel $N^{*}$ excitation amplitude
performed in our approach is particularly important for ISI \& FSI
treatment in $N^{*}$ excitation region.
For W above 2.0 GeV the contribution of resonant part in $\pi N$ elastic
scattering partial waves (Fig. 5) falls down drastically and non -- resonant
processes provide the main contribution for most partial waves.
Therefore a diffractive approximation for ISI and FSI absorptive
corrections \cite{Got64,LS71,Mur96} could be justified above $N^{*}$
excitation region.

To check the reliability of our approach
to evaluate ISI and
FSI absorptive factors, we also performed calculations of
such effects implementing \(\pi^{-}\Delta^{++}\) and \(\rho p\) elastic
scattering amplitudes directly provided
by the authors of analysis \cite{Dyt97,Dyt_unp}, 
based on a global unitary fit of $\pi N$ scattering
T-matrix. The gross features of ISI and FSI absorptive
factors as obtained from these two procedures
are in reasonable coincidence \cite{BatMSU,Bat98}.
It is also worth to note that using the above mentioned
description of Born terms, together with our ISI-FSI description,
non-resonant processes in our approach do not have any free parameters
to be determined from reaction (1).

Actually, our approach for ISI \& FSI treatment is mostly
phenomenological and our assumptions can be mainly justified
by comparison with experimental data (we will
discuss it in more detail in the next sections): pure Born terms
give rise to a cross section that does not reproduce real photon
data at high W by a large factor, while absorption-corrected
calculations are able to give a good account of data up to W
around 2 GeV.
Implementation of pion Regge trajectory exchange according to prescription
\cite{Gui97} in Born terms allows to obtain a satisfactory
data description up to W=3 GeV as discussed in section 4.

Assuming vector dominance as main mechanism responsible for \(Q^{2}\)
evolution of coupling with other hadronic channels in the initial state
we obtained the following expression for the
\(f_{ISI}^{j}\) absorptive factor \(Q^{2}\) dependence:
\begin{equation}
f_{ISI}^{j}(Q^{2}) = \frac{\Lambda_{\pi}^{2}f_{ISI}^{j}(Q^{2}=0)+Q^{2}}
{\Lambda_{\pi}^{2}+Q^{2}}
\end{equation}
where $\Lambda_{\pi}^{2}=0.46$ GeV$^{2}$\cite{Beb78}.

\subsection{Resonance contribution.}

A simple Breit-Wigner ansatz was chosen to describe the coherent
superposition of all relevant \(N^{*}\), \(\Delta^{*}\) resonant
amplitudes\cite{Pil79} (see Table I, Fig. 1a).
\begin{eqnarray}
\langle \lambda_{\Delta} \left| T_{res} \right| \lambda_{\gamma}\lambda_{p}
\rangle = \nonumber \\
\sum_{N^{*},\Delta^{*}}\langle \pi \lambda_{\Delta} \left| T_{dec} \right|
\lambda_{R} \rangle \frac{1}{M_{res}^{2}-W^{2}-i\Gamma_{res}(W)M_{res}}
\langle \lambda_{R} \left| T_{em} \right| \lambda_{\gamma}\lambda_{p} \rangle
\end{eqnarray}
where  \(M_{res}\), \(\Gamma_{res}\) are resonance mass and 
energy-dependent total width and
\(\langle \lambda_{R} \left| T_{em} \right|\lambda_{\gamma}\lambda_{p}\rangle\),
\(\langle \pi \lambda_{\Delta} \left| T_{dec} \right| \lambda_{R} \rangle\)
are electromagnetic production and strong decay
amplitudes of \(N^{*}\) with helicity
\(\lambda_{R}=\lambda_{\gamma}-\lambda_{p}\), respectively.
The \(N^{*}\) off-shell effects were taken into account
by the Breit-Wigner  propagator in (23) as well as by the W-dependence of
resonance width and strong decay amplitudes.

Following a phenomenological approach we related
\(\langle \lambda_{R} \left|
T_{em} \right| \lambda_{\gamma}\lambda_{p} \rangle\) and
\(\langle \pi \lambda_{\Delta}
\left| T_{dec} \right| \lambda_{R} \rangle\) amplitudes with
observables  extracted from experimental data  analysis.
Electromagnetic \(N^{*}\) amplitudes
\(\langle \lambda_{R} \left| T_{em}
\right| \lambda_{\gamma}\lambda_{p} \rangle\) were expressed  in  terms
of commonly used helicity couplings \(A_{1/2}\),
\(A_{3/2}\)
and \(C_{1/2}\)\cite{PDG96,Gia90}. Comparison of
the cross section for only one isolated \(N^{*}\) state calculated
according to (3)-(6) with Breit-Wigner formula gives:
\begin{eqnarray}
\langle \lambda_{R} \left| T_{em} \right| \lambda_{\gamma}\lambda_{p} \rangle
&=&
\frac{W}{M_{res}}\sqrt{\frac{8M_{N}M_{res}p_{\gamma_{R}}^{*}}{4\pi\alpha}}
\sqrt{\frac{p_{\gamma_{R}}^{*}}{p_{\gamma}^{*}}}\, A_{1/2,3/2}(Q^{2});\,
\left| \lambda_{\gamma}-\lambda_{p} \right| = \frac{1}{2},\, \frac{3}{2}
\nonumber \\
& & -for\, transverse\, photons
\nonumber \\
\langle \lambda_{R} \left| T_{em} \right| \lambda_{\gamma}\lambda_{p} \rangle
&=&
\frac{W}{M_{res}}\sqrt{\frac{8M_{N}M_{res}p_{\gamma_{R}}^{*}}{4\pi\alpha}}
\sqrt{\frac{p_{\gamma_{R}}^{*}}{p_{\gamma}^{*}}}\, C_{1/2}(Q^{2})
\\
& & -for\, longitudinal\, photons
\nonumber
\end{eqnarray}
where \(p_{\gamma_{R}}^{*}\) and \(p_{\gamma}^{*}\) are  photon three-momentum
moduli in the CM  frame  at
resonance point  (\(W=M_{res}\))  and at running W, respectively.
Couplings \(A_{1/2}(Q^{2})\), \(A_{3/2}(Q^{2})\), \(C_{1/2}(Q^{2})\)
completely describe \(N^{*}\) electromagnetic excitation and
their values, calculated in a model or taken from some experimental
analysis, can be used to
calculate the resonant part of cross section for reaction (1). On the other
hand, considering these couplings as
parameters our approach could be used to attempt their extraction
from a fit of the measured  cross section.

The \(N^{*}\) strong decay amplitudes
\(\langle \pi \lambda_{\Delta} \left| T_{dec} \right| \lambda_{R} \rangle\)
were parametrised through the projection on the set of states with
definite total angular momentum  j:
\begin{equation}
\langle \pi :\lambda_{\Delta} \left| T_{dec} \right| \lambda_{R} \rangle =
a_{\lambda_{\Delta}}^{j}d_{\lambda_{R}-\lambda_{\Delta}}^{j}(\theta^{*})
\sqrt{\frac{p_{\pi_{R}}^{*}}{p_{\pi}^{*}}}e^{-\lambda_{\Delta}i\varphi}
\end{equation}
where \(a_{\lambda_{\Delta}}^{j}\) is the decomposition coefficient,
that does not depend on the resonance helicity state due to rotational
invariance,
\(\theta^{*}\) is the CM pion emission angle,
\(p_{\pi_{R}}^{*}\)  and \(p_{\pi}^{*}\) are CM pion
three-momentum moduli at resonance and running W, respectively.
We  then related the \(a_{\lambda_{\Delta}}^{j}\) parameters
with  the partial decay widths \(\sqrt{\Gamma_{ls}}\) of
\(N^{*} \rightarrow \Delta^{++} + \pi^{-}\)  in   LS-
representation extracted in analysis \cite{Man92}.

This way we obtained the decay amplitudes at the resonance point.
General requirements for the amplitude threshold
behaviour\cite{New69} as well as data analysis of single-pion production
by photon and pion beams\cite{Wal69,Lon77} suggest the introduction of a
W dependence of \(N^{*}\)  decay
amplitudes in (27). Assuming as usually that barrier
penetration effects provide a good description of \(W\)- evolution
in \(\langle\pi\Delta (ls)|T_{dec}|R\rangle\) decay amplitudes and using
the parametrization from \cite{Man92} we arrived to:
\begin{eqnarray}
\langle \pi\Delta :(ls) \left| T_{dec} \right| R \rangle (W) =
\nonumber \\
\langle \pi\Delta :(ls) \left| T_{dec} \right| R \rangle (M_{res})
\left[ \frac{M_{res}}{W}
\frac{J_{l}^{2}(p_{\pi_{R}}^{*}R) + N_{l}^{2}(p_{\pi_{R}}^{*}R)}
{J_{l}^{2}(p_{\pi}^{*}R) + N_{l}^{2}(p_{\pi}^{*}R)} \right]^{1/2}
\end{eqnarray}
where \(J_{l}\), \(N_{l}\) are  Bessel's and Neumann's functions
and the  factor  in
square  brackets  in (30) appears from barrier penetration ratio\cite{Bla52}
for a particle emitted with relative orbital momentum \(l\), while  R
is an interaction radius whose value was set to 1 fm. The decay amplitude (30)
was then transformed  into the helicity representation
\(\langle \lambda_{\Delta} \left| T_{dec} \right|  \lambda_{R} \rangle\)
and using the relationship between two-body decay amplitude
and width \(\Gamma_{\lambda_{\Delta}}\) we obtained:
\begin{eqnarray}
\left| a_{\lambda_{\Delta}}^{j} \right| &=&
\frac{2\sqrt{2\pi}M_{res}\sqrt{2j+1}\sqrt{\Gamma_{\lambda_{\Delta}}}}
{\sqrt{\langle p_{\pi} \rangle}}
\nonumber \\
\langle p_{\pi} \rangle &=&
\int_{(m_{\pi}+m_{N})^{2}}^{(W-m_{\pi})^{2}}
dM^{2}\frac{1}{\pi}\frac{M_{\Delta}\Gamma_{\Delta}}{(M^{2}-M_{\Delta}^{2})^{2}
+M_{\Delta}^{2}\Gamma_{\Delta}^{2}}p_{\pi}(M^{2})
\\
p_{\pi}(M^{2}) &=& \frac{W^{2}+m_{\pi}^{2}-M^{2}}{2W}
\nonumber
\end{eqnarray}
where the integration over running \(\Delta\) mass squared $M^{2}$
takes into account the unstable character of \(\Delta\) particle
and the quantity $\sqrt{\Gamma_{\lambda_{\Delta}}}$ contains the
above mentioned W evolution according to (30).

The total \(N^{*}\) decay width (\(\Gamma_{res}(W)\) in (27)) was assumed
to be a sum over all partial widths presented in \cite{Man92}. The W evolution
of each partial width \(\Gamma_{i}(W)\) was evaluated again based on a barrier
penetration ansatz:
\begin{equation}
\Gamma_{i}(W) = \Gamma_{i}(W=M_{res})
\frac{M_{res}}{W}
\frac{J_{l}^{2}(p_{\pi_{R}}^{*}R) + N_{l}^{2}(p_{\pi_{R}}^{*}R)}
{J_{l}^{2}(p_{\pi}^{*}R) + N_{l}^{2}(p_{\pi}^{*}R)}
\end{equation}
where \(p_{\pi}^{*}\) and \(p_{\pi_{R}}^{*}\) are three-momenta
moduli of meson at running W and at resonance point respectively.

The  total amplitude of  \(\gamma p \rightarrow \Delta^{++}\pi^{-}\)
process was then evaluated as  a
coherent superposition of resonant and total Born amplitude
\(f_{\lambda_{\gamma}\lambda_{p}\lambda_{\Delta}}^{B^{corr}}\) , corrected
for ingoing and outgoing channel absorption:
\begin{equation}
\langle \pi \lambda_{\Delta} \left| T \right| \lambda_{\gamma}\lambda_{p} 
\rangle =
\langle \pi \lambda_{\Delta} \left| T_{res} \right| \lambda_{\gamma}\lambda_{p}
\rangle
+ f_{\lambda_{\gamma}\lambda_{p}\lambda_{\Delta}}^{B^{corr}}
\end{equation}

\section{Results and discussion.}
Using the above described approach we performed a cross section
calculation  for reaction (1). The
resonances included in the evaluation are listed in Table I.
The relative contribution of a particular \(N^{*}\) in the cross section can
be  described by the factor of merit
\(\sqrt{\Gamma_{\gamma p}\Gamma_{\Delta\pi}}/\Gamma_{tot}\)\(^{1}\)
\footnotetext[1]{Proportional to peak amplitude for Breit-Wigner curve.}
also reported in Table I. All three-  and
four-star  resonances\cite{PDG96} with
\(\sqrt{\Gamma_{\gamma p}\Gamma_{\Delta\pi}}/\Gamma_{tot} > 0.1\%\)
were included.
As  follows
from Table I, $F_{15}(1680)$, $D_{33}(1700)$ and $F_{37}(1950)$ resonances
give
maximum contribution and reaction (1) presents a promising opportunity
for investigation of their structure.

As mentioned in sect. 3.1, we adopted two ways for \(\Delta\)-in-flight
term evaluation: a) from gauge invariance requirements;
b) neglecting this term due to proximity of \(\Delta\) mass
and \(\Lambda\) cut-off parameter. 

Neglecting \(\Delta\)-in-flight term we obtained a better data
description at high W value and pion emission angles.
The results are presented in Fig. 6 by solid lines. The data are 
reasonably reproduced in the overall W region.
A remarkable point is that our calculations predict cross sections at
\(\theta_{\pi}^{*} > 120^{0}\) and \(W > 1.8\) GeV lower than 1.2 $\mu$b/sr;
such strong cross section
suppression namely results from absorption in the initial and final
states due to
interactions with open hadronic channels. For W = 1.62 GeV
the calculated cross sections are a bit below the data points;
this discrepancy could be ascribed to simultaneous uncertainties
in the cut-off parameter appearing in the strong vertex function used for
the Born terms, in the
\(\pi^{-}\Delta^{++}\rightarrow\pi^{-}\Delta^{++}\) ,
\(\rho^{0} p\rightarrow\rho^{0} p\)
elastic amplitudes appearing in the absorption
parameterisation, as well as to uncertainties in some s-channel \(N^{*}\)
electromagnetic and strong parameters.
The Born term contribution in angular distributions for reaction (1)
is shown by dashed lines in Fig. 6, therefore the
maximum $N^{*}$ contribution takes place for CM
pion emission angle above $90^{0}$. The evaluation of angular distributions
performed by varying $N^{*}$ strong decay couplings inside uncertainties 
of analysis \cite{Man92} demonstrated that the cross section variation does not
exceed a few percent, being negligible compared with data uncertainties:
therefore, 
our approach for the non-resonant background 
appears to be quite stable with respect to $N^{*}$ strong
decay parameter variation; this is an important feature for the extraction
of $N^{*}$ electromagnetic form factors from measured cross sections.

The comparison between the calculated total cross section for reaction (1)
with recent SAPHIR data\cite{Kle96} as well as with old
ABBHHM Collaboration data\cite{Bubble6768} is presented in Fig. 7. 
Decomposition of the total cross-section for reaction (1) in resonant,
non-resonant processes and interference terms is also shown
in Fig. 7. The maximum contribution of $N^{*}$ (at level $20-30\%$) is
found at $W < 1.6$ GeV and decreases as W increases. The region
$W < 1.6$ GeV also corresponds to maximum contribution of interference
effects in coincidence with results presented in \cite{Gom94,Och97}.

To demonstrate the effects of $\pi N \Delta$ form factor implementation
as well as ISI and FSI absorptive corrections, we also reported in Fig. 8
calculation
results assuming: a) $\pi N \Delta$ form factor equal to unity and
absence of absorptive corrections (dashed line in Fig. 8);
b) $\pi N \Delta$
form factor from \cite{Mac87} and absence of
absorptive corrections (dotted
line in Fig. 8); c) complete model (solid line). Calculation a)
is not able to reproduce data at
all: at $W > 1.6$ GeV the difference between calculated and measured
cross-section marks the complete 
failure of the pure Born terms calculation.
Implementation of $\pi N \Delta$ strong
form factor turns out to be very important (dotted line of Fig. 8):
we stress that such strong vertex function was taken from N -- N
scattering analysis without further tuning and assumed to be
appropriate to describe the $\pi N \Delta$ vertex in the t-channel
pion exchange diagram, while for the contact term which is of
similar magnitude as the pion exchange, 
we used gauge invariance as guidance for the vertex functions
evaluation.
However without initial and final state absorptive
corrections it is not possible to reproduce data at W above 1.7 GeV.
On the other hand, even neglecting ISI and FSI effects our approach
is able
to describe reasonably the data at $W < 1.6$ GeV.
Implementation of ISI and FSI absorptive correction factors
provides a reasonable agreement between
measured and calculated cross sections
for W below 1.8 GeV, while for W above 1.9 GeV calculated cross sections
are systematically higher than data. The reason for such
deviation could be connected to
the description of \(\pi\Delta\),
\(\rho p\) elastic scattering amplitudes in ISI-FSI: this aspect
could be improved by
adding
contributions from isobar states with higher mass and spin;
another explanation
could be that for W above 1.9 GeV we may start to observe a transition
from the corrected Born terms picture to other processes,
as those discussed in \cite{Mur96,Gui97};
in this case, the framework for non-resonant reaction mechanisms
in \(\gamma p\rightarrow\pi^{-}\Delta^{++}\) reaction should be
modified at W above 1.9 GeV. We discuss these aspects in the next 
paragraph, to examine the limits of applicability of our meson-baryon
approach and study how to continue the description of reaction (1)
in the higher energy region.

Following the recipe of \cite{Mur96,Gui97}, we replaced the pion exchange
amplitude in Born term (Fig. 1c) by a pion Regge trajectory exchange.
The results are presented in Fig. 9 (dotted line) and compared with 
the non-reggeized pion exchange (solid line). Regge trajectory
exchange implementation provides a better data description at high W values.
However, even in this case the calculated
cross section appears to be systematically higher than the data. 
The possible reason could be the need to modify also the contact term in
the way proposed in \cite{Gui97}, thereby restoring gauge invariance.
We found that such contact term modification led to a sizeable cross section
reduction at W below 1.6 GeV, where descriptions in the frame of pion and
Regge trajectory exchange have to coincide. To provide this coincidence we
had to require $N \Delta$ -- (pion Regge trajectory) effective coupling
to be a factor 1.2 higher than $\pi N \Delta$ coupling. The cross section 
calculated under
this assumption is shown in Fig. 9 with a dashed 
line and reproduces reasonably the data up to 2.5 GeV. As mentioned 
above, non -- resonant part for $\pi^{-} \Delta^{++}$ and $\rho p$ elastic
scattering amplitudes in ISI-FSI was estimated from $\pi N$ scattering
partial wave data fit\cite{Man92},
containing data up to W  $=$ 2.1 GeV. This is the main reason to restrict our
calculation to W $=$ 2.5 GeV (3 GeV photon energy in lab. frame).
Implementation of any data on
$\pi^{-} \Delta^{++}$ and $\rho p$ elastic scattering amplitudes above
2 GeV would allow a model extension toward higher W.

To investigate the influence of $\gamma p N^{*}$ vertex dressing on our
cross section, we reported in Fig. 10 the calculations
with $N^{*}$ electromagnetic helicity amplitudes taken from PDG\cite{PDG96}
and from quark models\cite{Kon80,Bij96,Gia98}.
In the PDG case, we assumed
the effects of $\gamma p N^{*}$ vertex dressing at level of meson --
baryon degrees of freedom to be included in the numbers quoted, 
while in the quark models case we assumed
such effects not to be accounted for ("bare" $\gamma p N^{*}$ 
vertices). 
Therefore, using the PDG amplitudes we applied our prescription
(24) for removal of $N^{*}$ dressing in ISI-FSI, while
in cross section calculations with $\gamma p N^{*}$
vertices from quark models we kept $N^{*}$ excitation terms in
ISI \& FSI mechanisms with no modification. 
In principle, one would expect these two results approximately
to coincide, assuming that the same dressing effects are 
introduced one way or another. Our results show that this
is not the case, being the cross section calculated using 
the "dressed" PDG values systematically higher than
the quark model results; moreover, if we removed completely
the $N^{*}$ from the ISI-FSI description, the discrepancy
would be even bigger; we also found that the difference
between the two
quark model cross sections is significantly smaller 
than the deviation between quark model and PDG results.
Actually, the difference between the cross section evaluated with
"bare" and "dressed" vertices could be due to specific approximations
of quark model approaches\cite{Kon80,Bij96,Gia98};
or it could be an indication that additional
dressing effects are present in the experimentally extracted
photocouplings from \cite{PDG96}. 
This difference 
does not exceed 30 \%, with a maximum at W between 1.5 -- 1.7 GeV. 
This W range
actually corresponds to maximum $N^{*}$ contribution. 
The discrepancy appears
to be negligible at W above 2.0 GeV and below 1.4 GeV, where $N^{*}$ 
contributions are less pronunced.

To see the influence of particular dressing effects -- $N^{*}$ excitation
in $\pi^{-} \Delta^{++}$ and $\rho p$ elastic scattering processes shown
in Fig. 4c -- we performed calculations using
"dressed" $N^{*}$ electromagnetic vertices\cite{PDG96}, 
but keeping the $N^{*}$ excitation
in ISI \& FSI mechanisms for Born terms.
The results are shown in Fig. 10 by dotted line; of course 
in this case some double counting of the $N^{*}$ electromagnetic
vertex dressing takes place, but surprisingly
the cross section seems to be in better agreement with
the "bare" quark model results.
In any case, 
the difference between solid and dotted curves could be considered as
estimation of this particular $\gamma p N^{*}$ vertex dressing contribution 
in the cross section. This contribution is lower than 20 \% and vanishes
at W above 2.0 GeV, where $N^{*}$ excitation is negligible. 

To extract $N^{*}$ helicity amplitudes for $Q^{2} > 0$ it is important
to have a good description of the $Q^{2}$ dependence
for non-resonant processes.
We checked this point by comparing the total cross section
for reaction (1)
calculated in our approach with data reported in \cite{Wac78}.
For the resonant part
we used results from a Single Quark Transition Model (SQTM) 
fit\cite{Bur94}
about the $Q^{2}$ evolution of $N^{*}$ photocouplings. The
comparison between data\cite{Wac78} and our calculations is
presented in Fig. 11  where dashed lines correspond to the
contribution of non -- resonant
processes only, while the complete evaluation with $N^{*}$
included is shown by solid lines.
Considering this comparison, two important points must be stressed:
first, bins for experimental data\cite{Wac78}, both in W and $Q^{2}$
were very
wide, implying a strong cross section averaging over the measured
kinematical range; second, experimental
data about high-lying $N^{*}$ photocouplings are rather scarce  and
do not allow to check the validity of SQTM predictions.
However, our approach reproduces
the measured cross-section for W in the 1.3-1.5 GeV and
in the  1.5-1.7 GeV interval; for W in the  1.7-2.0 GeV bin
our calculated cross-section is  higher than the measurements;
the reasons for
such deviation could be namely the poor knowledge of the
$N^{*}$ contributions or the kinematical average effects,
as well as the above mentioned limitations in the treatment
of Born terms absorption in the higher W region.
New high precision data upcoming from new facilities
like TJNAF are therefore definitely needed for a better understanding
of the \(Q^{2}\) behaviour of $N^{*}$ electromagnetic form factors.
An important prediction of our evaluations is a strong $N^{*}$
contribution in the W = 1.5 -- 1.7 GeV interval,
representing over 60 \% of cross section
at $Q^{2} \sim 1.0$ GeV$^{2}$. 
Therefore the two pion exclusive channel seems
to present a promising opportunity for $N^{*}$ structure investigation by
photons with high virtuality.

To investigate $N^{*}$ electromagnetic vertex dressing effects on the
cross section at $Q^{2} > 0$, we performed the calculation 
with $N^{*}$ electromagnetic form factors taken from different
approaches\cite{Bij96,Gia98,Bur94}. 
Form factors in approach\cite{Bur94} were determined from exclusive
single pion production data analysis, imposing symmetry relations between
form factors of $N^{*}$'s belonging to a particular SU(6) multiplet: therefore
they can be assumed to represent "dressed" vertices. Instead, 
quark models \cite{Bij96,Gia98} should not contain higher order corrections 
and we considered their results as representing "bare" vertices. 
Again we expected to have approximate coincidence of results
obtained using these different ingredients. Actually in this case, 
according to Fig. 12, the difference between cross sections
calculated with the two "bare" vertices \cite{Bij96} 
and \cite{Gia98} is higher than the difference
between results obtained using form factors from \cite{Bur94}
and quark models results. Also in this case we calculated
the cross section using "dressed" vertices from \cite{Bur94},
with and without prescription (24) for  $N^{*}$ removal
in ISI-FSI (solid and dotted curve in Fig. 12, respectively);
in the latter case the result is supposed to be affected by
double counting, but the difference can give an indication
of the extent of dressing introduced by our effective
unitarity implementation; as shown in Fig. 12, such effect
appears to be negligible, thereby indicating that, for the
$Q^{2} > 0$ evolution in our framework, quark model
theoretical uncertainties could be more important than
dressing effects.
Therefore, comparison
of cross sections calculated in our approach with forthcoming 
precise experimental data at $Q^{2} > 0$ can provide a 
promising opportunity to select between model approaches for $N^{*}$ 
structure description in non -- perturbative QCD region.

As mentioned above, we also compared the cross section calculated using
"dressed" $N^{*}$ form factors\cite{Bur94} with/without 
implementation of $N^{*}$
excitation in ISI \& FSI mechanisms
(dotted and solid lines in Fig. 12). The contribution
of this particular dressing mechanism is negligible for 
W$\sim$ 1.4 and 1.85 GeV
and vanishes at $Q^{2} > 0.5$ GeV$^{2}$ for W $\sim 1.6$ GeV.
There are actually reasons for such behaviour:
a) the relative contribution of non -- resonant processes drastically 
falls down as $Q^{2}$ increases (from more than 80 \% at the photon point
to lower than 50 \% at $Q^{2}$ above 1 GeV$^{2}$; b) the ISI effects also
go down as $Q^{2}$ increases due to suppression of transitions between
photon and vector meson.

\section{Conclusions.}
We  have  extensively studied a phenomenological model for the two
pion  photo- and electroproduction through the intermediate \(\Delta^{++}\pi^{-
}\)
channel, as part
of  a  broader  effort  in establishing a basis for analysis and interpretation
of the upcoming data from TJNAF\cite{Bur98,Rip98}, 
where this  as  well  as  other
exclusive  electromagnetic production channels  will  be  studied
with unprecedented accuracy.

Our  calculation included a minimal set of Born  amplitudes,
with  appropriate absorptive corrections  to
effectively take into
account interaction with open channels in the initial and final states,
plus a  large  number  of
nucleon  resonances  that, according to the  existing  data,  are
thought to give a sizeable contribution to the cross section.

A strong form factor for \(\pi N\Delta\) vertex was
introduced according to NN scattering experiments.
Data about pion and proton
electromagnetic form factors were also used to
evaluate the pion-in-flight term and the  s-channel nucleon
pole term, respectively, for \(Q^{2}>0\), while the \(Q^{2}\)
behaviour of the contact term was obtained
imposing gauge invariance;  the delta-in-flight term was calculated
both from gauge invariance and using a \(\pi N\Delta\) vertex
determined from NN scattering analysis.
Strong and electromagnetic \(N^{*}\) couplings
were related with experimental observables: electromagnetic
helicity couplings \(A_{1/2}\), \(A_{3/2}\), \(C_{1/2}\) and
partial hadronic decay widths \(\Gamma_{ls}\).

We found that strong absorptive corrections were essential to
reproduce experimental data for CM pion emission angle above 30\(^{0}\)
 at W above 1.6 GeV.
We developed a specific approach for ISI and FSI absorptive corrections,
relating absorptive factors in ingoing
and outgoing channels with \(\pi^{-}\Delta^{++}\) and \(\rho p\)
elastic scattering amplitudes. These elastic strong amplitudes were evaluated
in a simple isobaric model, so that our approach did not have any free
parameters to
be determined from reaction (1).
Any other approach for \(\pi^{-}\Delta^{++}\) and \(\rho p\) elastic
amplitude evaluation can be easily implemented in our calculation.
Reasonable reproduction of pion angular distributions in the photon
point as well as good agreement between measured and calculated total
cross sections for \(Q^{2} > 0\) demonstrate the ability of our
approach to reproduce the main features of \(\pi^{-}\Delta^{++}\)
production by real and virtual photons in the resonance region.
Moreover the proposed method
for ISI and FSI description can be considered as a reasonable phenomenological
way to describe non-resonant processes in electromagnetic meson production
for \(W > 1.6\) GeV, where the competition of many open hadronic
channels makes a rigorous background evaluation very difficult.
Our unitarity corrections effectively introduce a dressing
of the resonance amplitudes; therefore we elaborated
an empirical prescription to remove the dressing
from ISI-FSI when using experimental resonance photocouplings,
assumed to be already ``dressed''; we also
performed calculations without this prescription
when using resonance amplitudes from quark
models, asssumed to be ``bare''. In the real photon
case, we found that the calculations performed
with two different quark models show a good 
agreement, while the results obtained using
the experimental PDG amplitudes show a systematic
disagreement with the previous ones; the reason
is unclear and could be connected to other dressing 
effects than $N^{*}$ excitation in 
$\rho p \rightarrow \rho^{'} p^{'}$
$\pi^{-'} \Delta^{++'} \rightarrow \pi^{-} \Delta^{++}$ 
transitions
as well as to quark model uncertainties and also to 
model-dependence
in the experimental resonance amplitude extraction.
We performed the same analysis for the virtual
photon cross section, but in this case we found
a disagreement between the two quark models
adopted, while the calculation based on resonance
amplitudes extracted from an analysis of experimental 
data showed less sensitivity to our ``dressing''
effects: the quark models discrepancies appeared
to be more important, leading us to believe that
in the virtual photon case the sensitivity to
different quark model ingredients could be more
pronunced, although the reason for the different
behaviour of real and virtual photon is not evident.

The  calculation presented
could   serve  as  a  first  basis   for
interpreting the data coming from new facilities like Jefferson Lab and
moreover, allowing the resonance parameters to
vary, it could be the starting point of a fitting procedure  with
the  goal  of investigating electromagnetic form factors  for  high
mass  \(N^{*}\)'s (\(M_{res} > 1.5\) GeV), as well as for attempting
to discover  new
states. The model presented is currently being used as a foundation  in our
development of a full three-body final state description
for the  two-pion
production off the nucleon by real and virtual photons\cite{Rip98,Mok98}.

\vskip10pt
{\bf Acknowledgements}
\vskip10pt
Our special thanks to Dr. V. Burkert from Jefferson Laboratory,
for continuos interest, useful discussions and support. 
Particular thanks to
Prof. S. Dytman from the University
of Pittsburgh, for kindly providing
the elastic hadronic amplitudes from his global fit
and to Prof. M. Giannini from the
University of Genova, Italy, for kindly
providing the \(N^{*}\) electromagnetic transition
amplitudes from his quark model.
We also want to thank Prof. N.C. Mukhopadhyay from
Rensselaer Polytechnic Institute, for useful discussions and
suggestions.

\newpage

\vskip10pt
{\bf Figure captions}
\vskip10pt

Fig 1. Tree-level diagrams for the \(\pi^{-}\Delta^{++}\)
electromagnetic production on
proton. 
\vskip10pt

Fig2. ISI and FSI mechanisms.
\vskip10pt

Fig3. Description of $\pi \Delta$ and $\rho N$ elastic
scattering amplitudes related to ISI and FSI.
\vskip10pt

Fig 4. The effective $\gamma p N^{*}$ vertex (a), 
the bare $\gamma p N^{*}$ vertex
(b), the dressing vertex correction, containing $N^{*}$ excitation in
$\pi^{-} \Delta^{++}$ and $\rho p$ elastic scattering.
\vskip10pt

Fig 5a. $\pi N$ partial wave cross section from
\cite{Man92} decomposed in resonant (dotted
lines) background (dashed lines) parts according to procedure described
in sect. 3.5. The fit results are shown by solid lines.
\vskip10pt

Fig 5b. (continued)
\vskip10pt

Fig 5c. (continued) 
\vskip10pt

Fig 6. The calculated and measured\cite{Bubble6768} 
angular distributions for
$\gamma p \rightarrow \Delta^{++}\pi^{-}$ reaction
assuming negligible contribution
for $\Delta$-in-
flight term (see text sect. 3). Dashed lines
correspond to Born terms alone, while solid lines
represent the complete calculation with $N^{*}$ contribution included.
\vskip10pt

Fig 7. Decomposition of
$\gamma p \rightarrow \Delta^{++}\pi^{-}$ reaction
cross section in Born terms (dashed line), $N^{*}$ terms (dotted line)
and interference term (dash-dotted line) contributions.
Data are from \cite{Bubble6768} (squares) and from
\cite{Kle96} (circles and triangles).
\vskip10pt

Fig 8. Total cross section for
$\gamma p \rightarrow \Delta^{++}\pi^{-}$ reaction
at the photon point. Dashed line represents 
calculation results with no $\pi N \Delta$
form factor and no ISI and FSI absorptive corrections.
Dotted line corresponds to $\pi N \Delta$ strong form factor
taken from \cite{Mac87} and
no ISI and FSI absorptive corrections. Solid line represents
cross section evaluation in the complete model.
Data as in Fig. 8.
\vskip10pt

Fig 9. The calculation with pion Regge trajectory exchange\cite{Gui97}.
Solid line represents calculations performed with Born terms evaluated in
single pion exchange picture. Dotted line represents the results obtained
after substitution of pion exchange by pion Regge trajectory exchange,
dashed line corresponds to an additional modification of contact term
according to prescription of \cite{Gui97}. Data as in Fig. 8.
\vskip10pt

Fig 10. The influence of $\gamma p N^{*}$ vertex dressing effects at
the photon point.  Our calculations with 
PDG\cite{PDG96} $\gamma p N^{*}$
vertices (``dressed'') 
and $N^{*}$ excitations excluded (solid line) and included
(dotted line) in ISI \& FSI treatment for Born terms are presented. 
The calculation results with $\gamma p N^{*}$ vertices
calculated in quark models ("bare" vertices) and with $N^{*}$
excitation included in ISI \& FSI are shown by:
dashed line for model \cite{Kon80}, dot -- dashed line for model 
\cite{Gia98}. Data as in Fig. 8.
\vskip10pt

Fig 11. $Q^{2}$ dependence of
$\gamma_{v} p \rightarrow \Delta^{++}\pi^{-}$ total
virtual photon
cross section in comparison with data from \cite{Wac78}.
\vskip10pt

Fig 12. $Q^{2}$ dependence of  $\gamma p N^{*}$ vertex dressing effects.
Our calculations with $\gamma p N^{*}$
vertices from \cite{Bur94} (``dressed'') 
and $N^{*}$ excitations excluded (solid line) and included
(dotted line) in ISI \& FSI treatment are presented. 
The calculation results with $\gamma p N^{*}$ vertices
from quark models ("bare" vertices) and with $N^{*}$
excitation included in ISI \& FSI treatment are shown by:
dashed line for model \cite{Bij96}, 
dot -- dashed line for model \cite{Gia98}. Data from \cite{Wac78}.
\vskip10pt

\newpage

\vskip10pt
{\bf Table captions}
\vskip10pt

Table 1. List of resonances included in our calculation.
\vskip10pt

Table 2. List of the resonances predicted by quark models that are
weakly coupled to the
\(N\pi\) channel but should be strongly coupled to the \(N\pi\pi\)
channels\cite{Kon80}.
\vskip10pt

Table 3. Nucleon resonances contributing in \(\pi^{-}\Delta^{++}\) and
\(\rho^{0} p\)
elastic scattering amplitudes.
\vskip10pt

\newpage

\begin{table}
\caption{}
\vspace{2mm}
\begin{tabular}{|c|c|c|c|c|c|} \hline
Resonance,         &  Mass,    &   Width,  &   \(\Gamma_{\Delta\pi}\),   &
\(\Gamma_{N\rho}\),  &  \(\frac{\sqrt{\Gamma_{\gamma
p}\Gamma_{\Delta\pi}}}{\Gamma_{tot}}\), \\
  parity           &  (MeV)    &   (MeV)  &           (\%)              &
(\%)          &       (\%)  \\ \hline
\(P_{11}(1440)\) + & 1430-1470 & 250-450    &        20 - 30            &
\(<\) 8         &            1.06\\ \hline
\(D_{13}(1520)\) - & 1515-1530 & 110-135    &        15 - 25            &
15 - 25       &             0.99\\ \hline
\(S_{11}(1650)\) - & 1640-1680 & 145-190    &         1 - 7             &
4 - 12       &             0.64\\ \hline
\(D_{15}(1675)\) - & 1670-1685 & 140-180    &        50 - 60            &
\(<\) 1-3        &             0.70\\ \hline
\(F_{15}(1680)\) + & 1675-1690 & 120-140    &        5 - 15             &
3 - 15        &             1.59\\ \hline
\(P_{13}(1720)\) + & 1650-1750 & 100-200    &                           &
70-85         &               \\ \hline
\(S_{31}(1620)\) - & 1615-1675 & 120-180    &        30 - 60            &
7 - 25        &             0.97\\ \hline
\(D_{33}(1700)\) - & 1670-1770 & 200-400    &         30 - 60           &
30 - 55        &            3.16\\ \hline
\(F_{35}(1905)\) + & 1870-1920 & 280-440    &      \(<\) 25             &
\(>\) 60         &            0.60\\ \hline
\(F_{37}(1950)\) + & 1940-1960 & 290-350    &         20 -30            &
\(<\) 10         &            1.50\\ \hline
\end{tabular}
\end{table}

\begin{table}
\vspace{15mm}
\caption{}
\vspace{2mm}
\begin{tabular}{|c|c|c|c|c|c|} \hline
Resonance,        &  Mass,  & \(\Gamma\), & \(\Gamma_{N\pi\pi}\), &
\(\Gamma_{\Delta\pi}\), & \(\Gamma_{N\rho}\), \\
 parity           &  (MeV)  &     (MeV)   &        (\%)           &     (\%)
&       (\%)   \\ \hline
  \(P_{13}\) +    &  1870   &      150    &         14            &      13
&         1   \\ \hline
  \(P_{11}\) +    &  1890   &      100    &         34            &      12
&        22   \\ \hline
  \(P_{13}\) +    &  1955   &      250    &         63            &      36
&        27   \\ \hline
  \(F_{15}\) +    &  1955   &      320    &         41            &      20
&        21   \\ \hline
  \(F_{35}\) +    &  1975   &      430    &        100            &       9
&        91   \\ \hline
  \(P_{33}\) +    &  1975   &       95    &         99            &      62
&        37   \\ \hline
  \(P_{13}\) +    &  1980   &      220    &         76            &      44
&        32   \\ \hline
  \(P_{11}\) +    &  2055   &      40     &         12            &       8
&        4   \\ \hline
  \(P_{13}\) +    &  2060   &      140    &         33            &      22
&        11   \\ \hline
\end{tabular}
\end{table}

\newpage
\begin{table}
\caption{}
\vspace{2mm}
\begin{tabular}{|c|c|} \hline
Resonance spin &
\(N^{*}\) included in resonant part of the amplitudes \\ \hline
1/2 &
S11(1535), S31(1620), S11(1650), \\
 &
P11(1440), P11(1710), P31(1910) \\ \hline
3/2 &
P13(1720), P33(1600), P33(1920), \\
 &
D13(1520), D13(1700), D33(1700) \\ \hline
5/2 &
D15(1675), F15(1680), F35(1905) \\ \hline
\end{tabular}
\end{table}

\newpage
\begin{figure}[h]
\vspace{2cm}
\begin{center}\epsfig{file=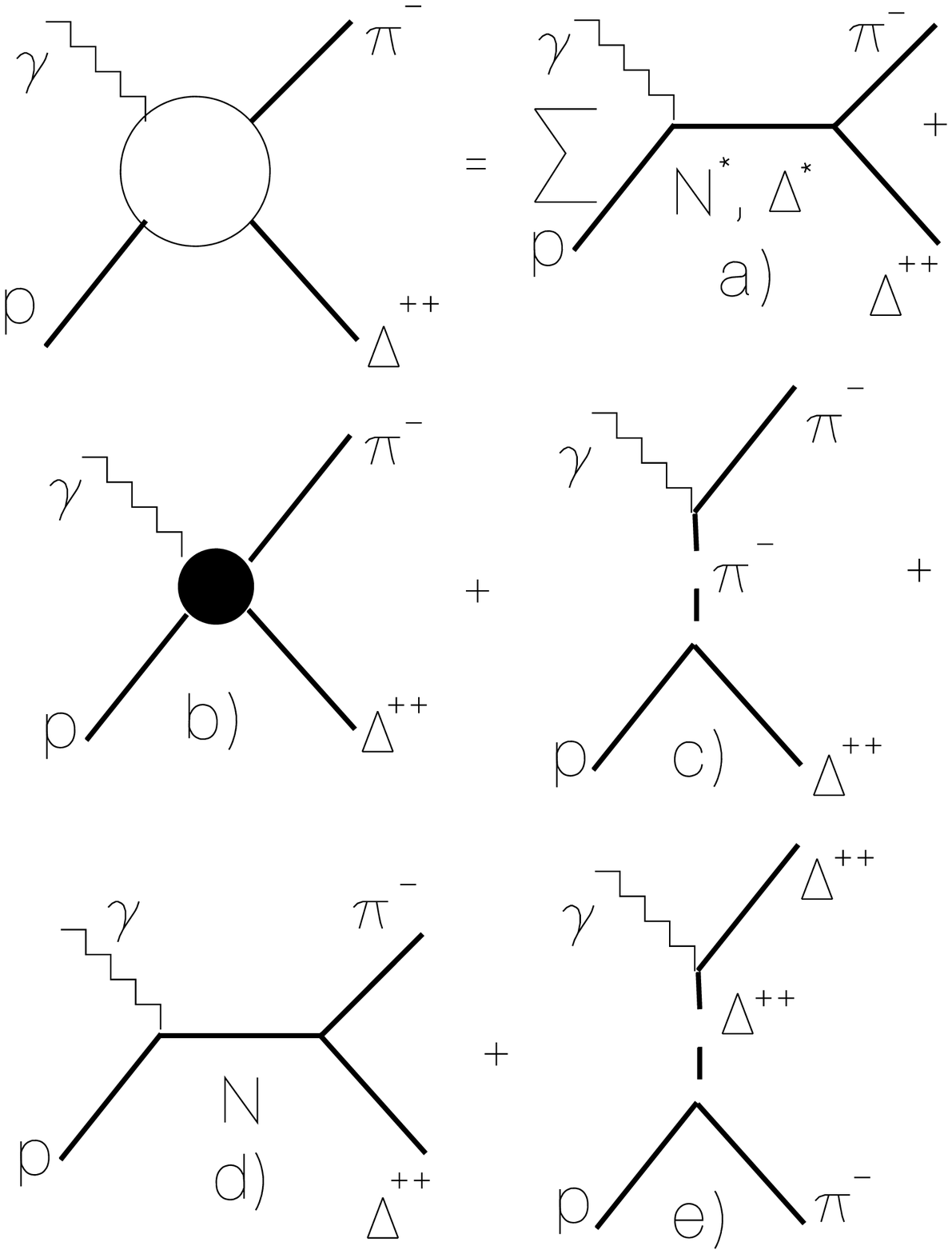,width=15cm}\end{center}
Figure 1
\end{figure}

\newpage
\begin{figure}[h]
\vspace{2cm}
\epsfig{file=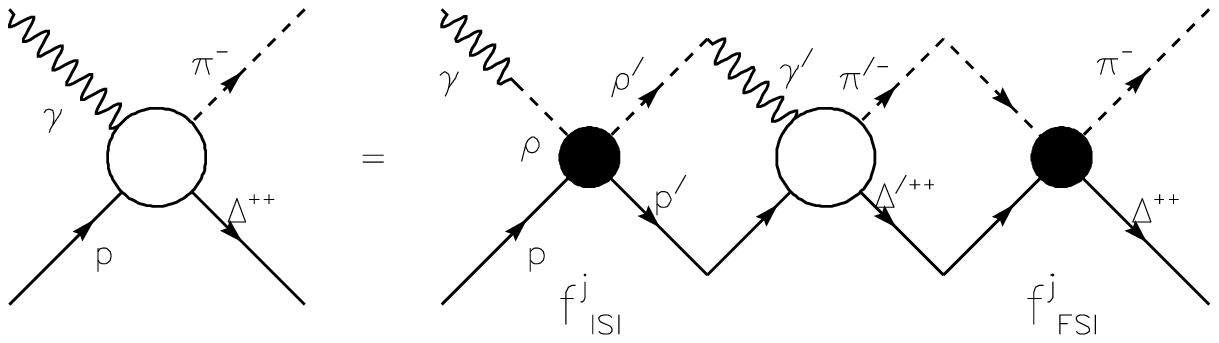,width=15cm}
\nonumber
Figure 2
\end{figure}

\newpage
\begin{figure}[h]
\begin{center}\epsfig{file=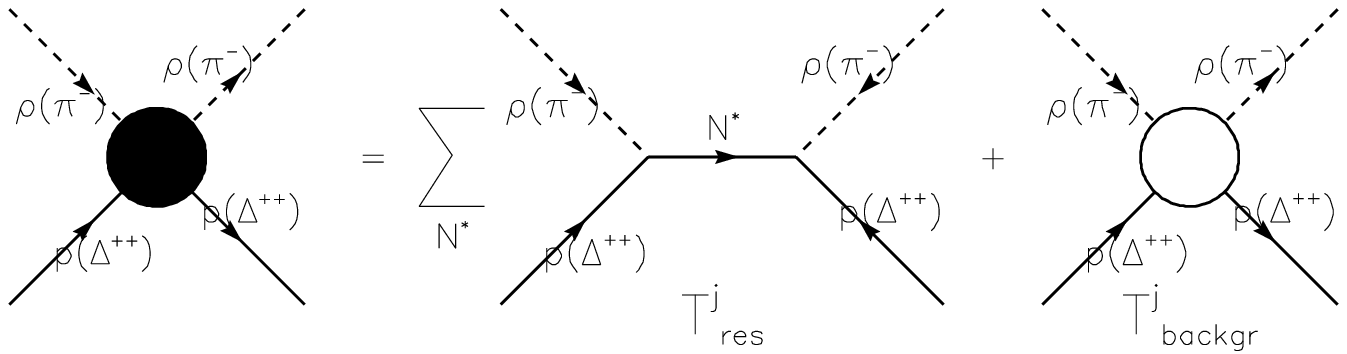,width=15cm}\end{center}
Figure 3
\end{figure}

\newpage
\begin{figure}[h]
\begin{center}\epsfig{file=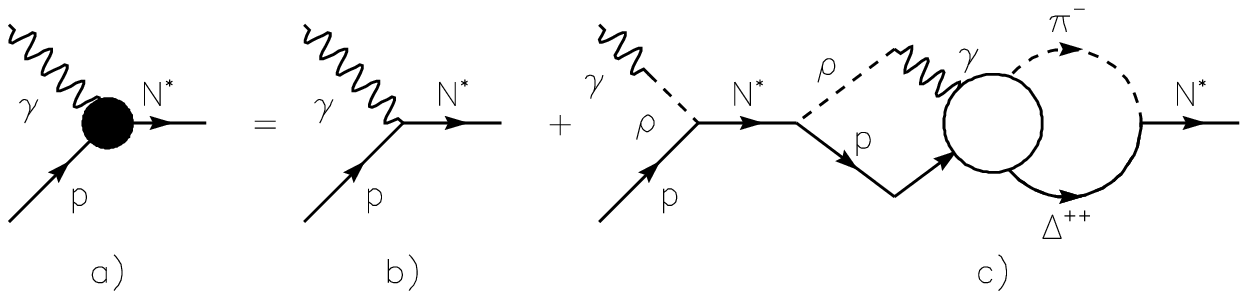,width=15cm}\end{center}
Figure 4
\end{figure}

\newpage
\begin{figure}[h]
\epsfig{file=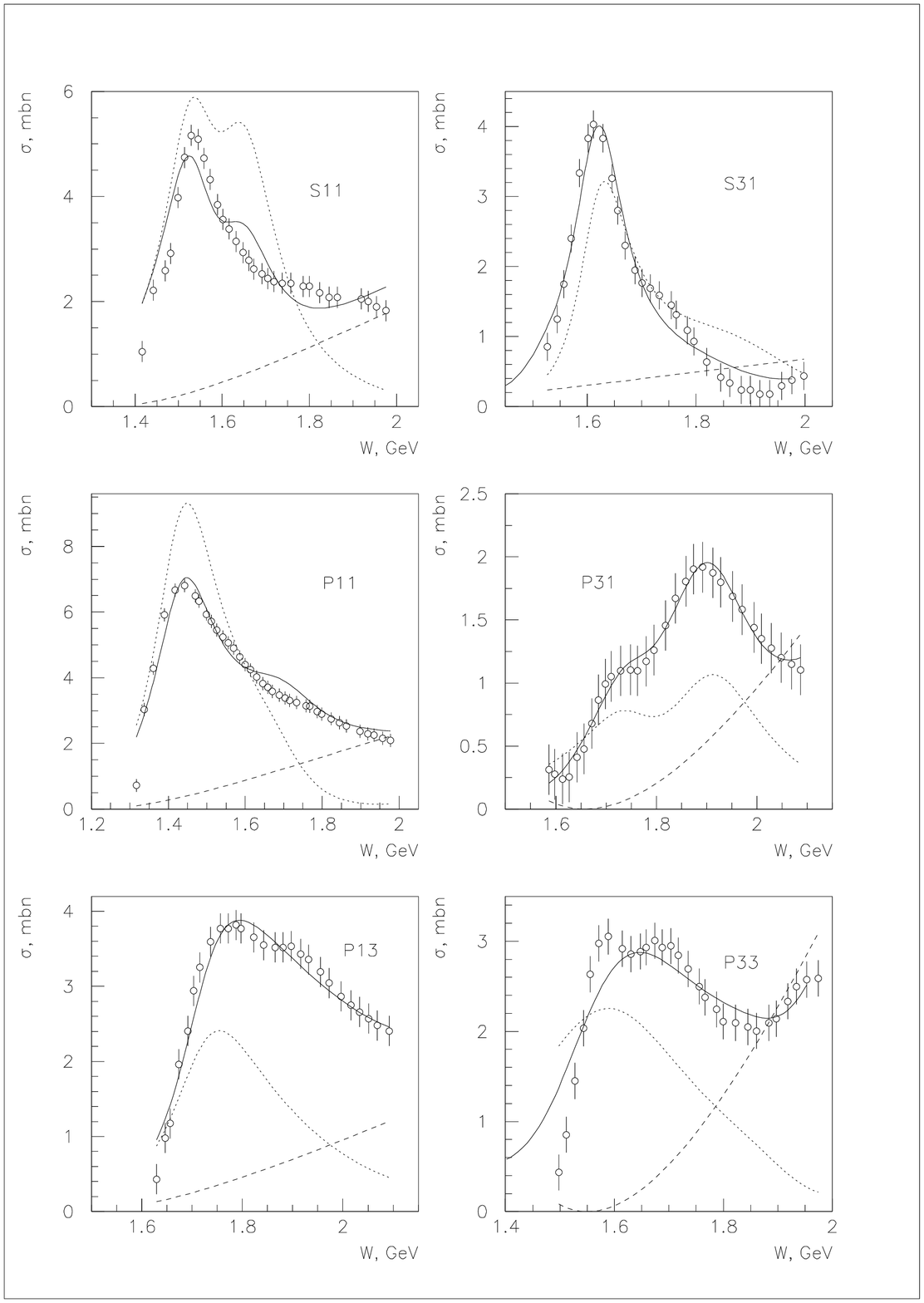,width=15cm}
\nonumber
Figure 5a
\end{figure}

\newpage
\begin{figure}[h]
\addtocounter{figure}{-1}
\epsfig{file=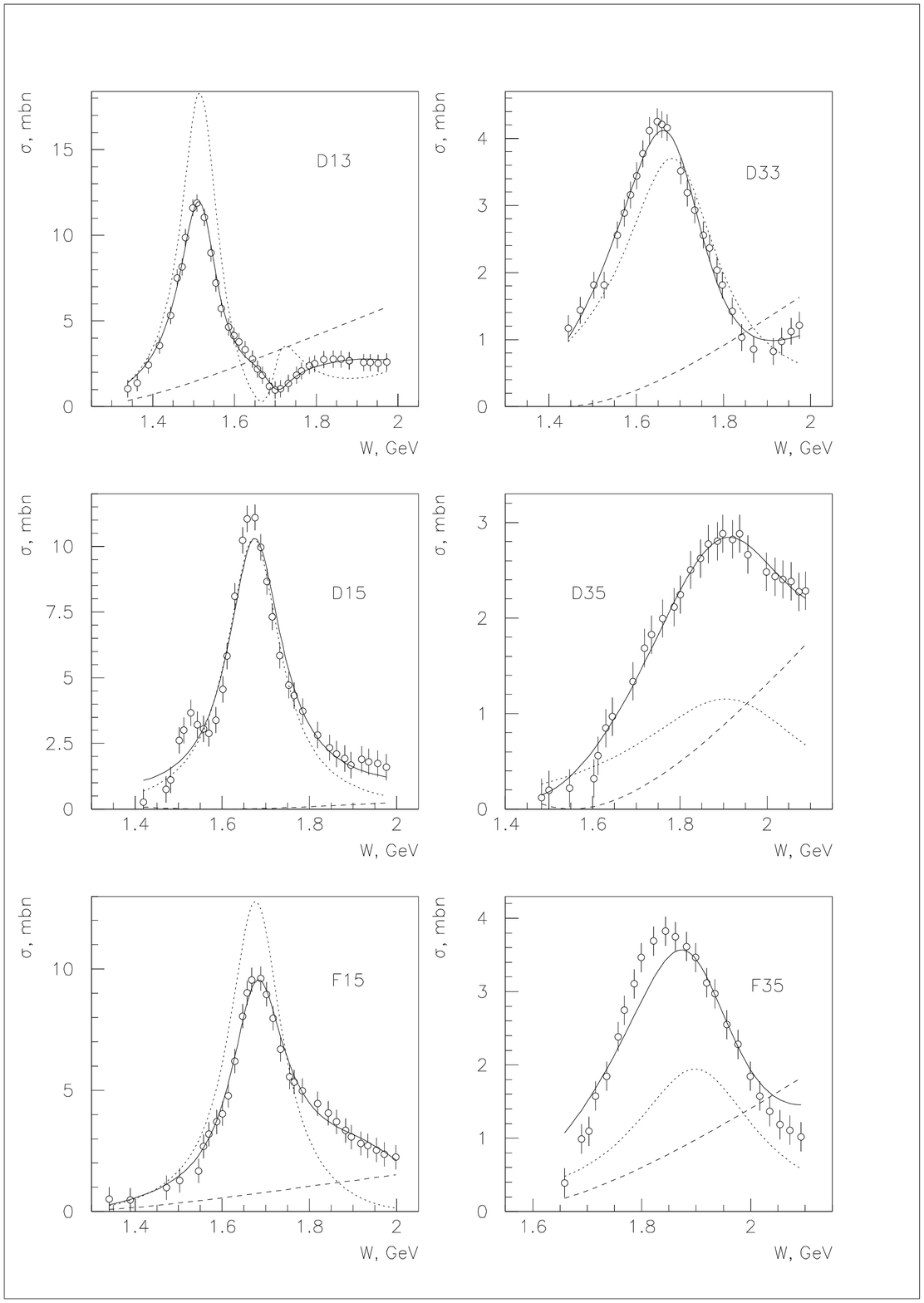,width=15cm}
Figure 5b
\end{figure}

\newpage
\begin{figure}[h]
\addtocounter{figure}{-1}
\epsfig{file=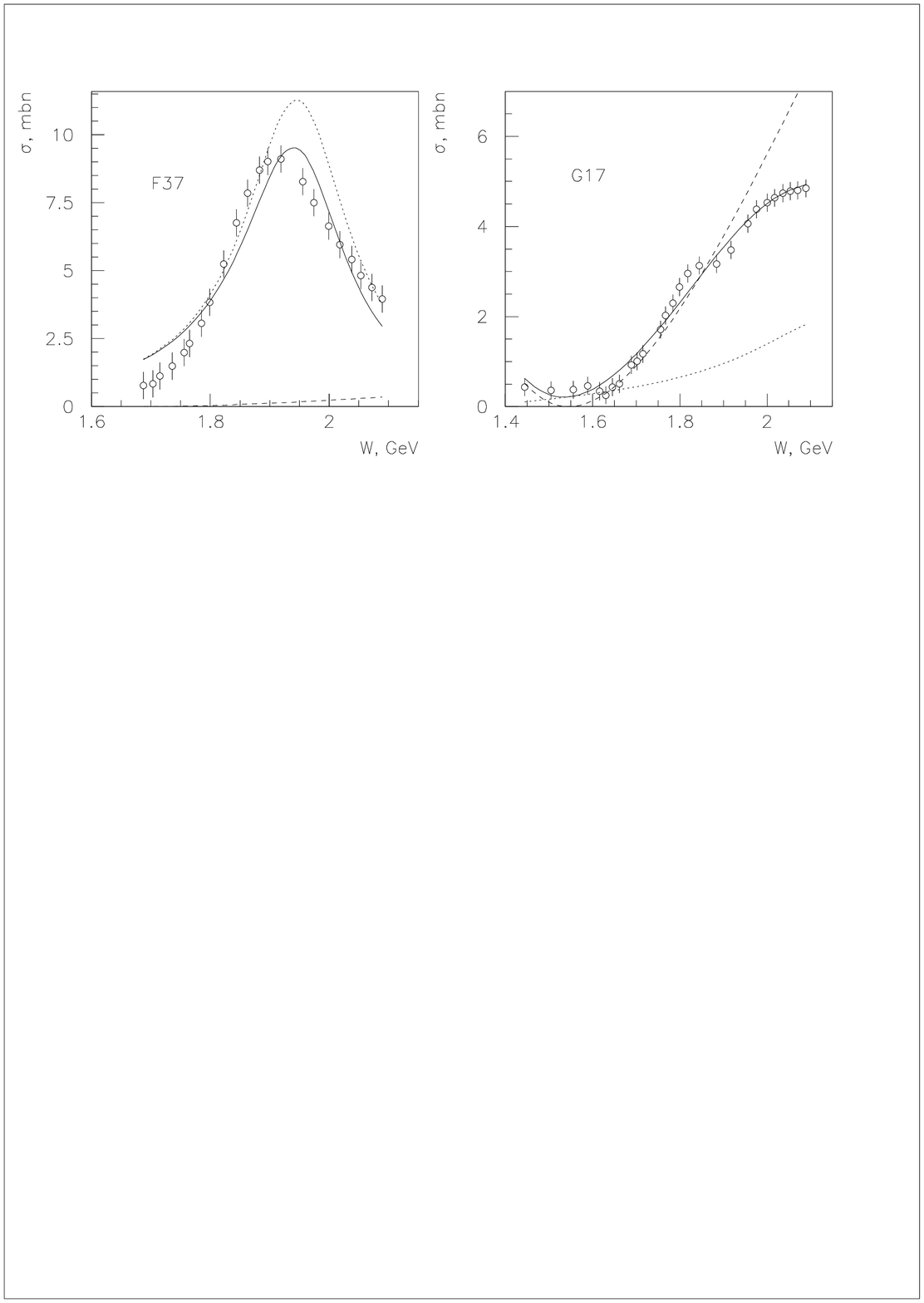,width=15cm}
Figure 5c
\end{figure}

\newpage
\begin{figure}[h]
\epsfig{file=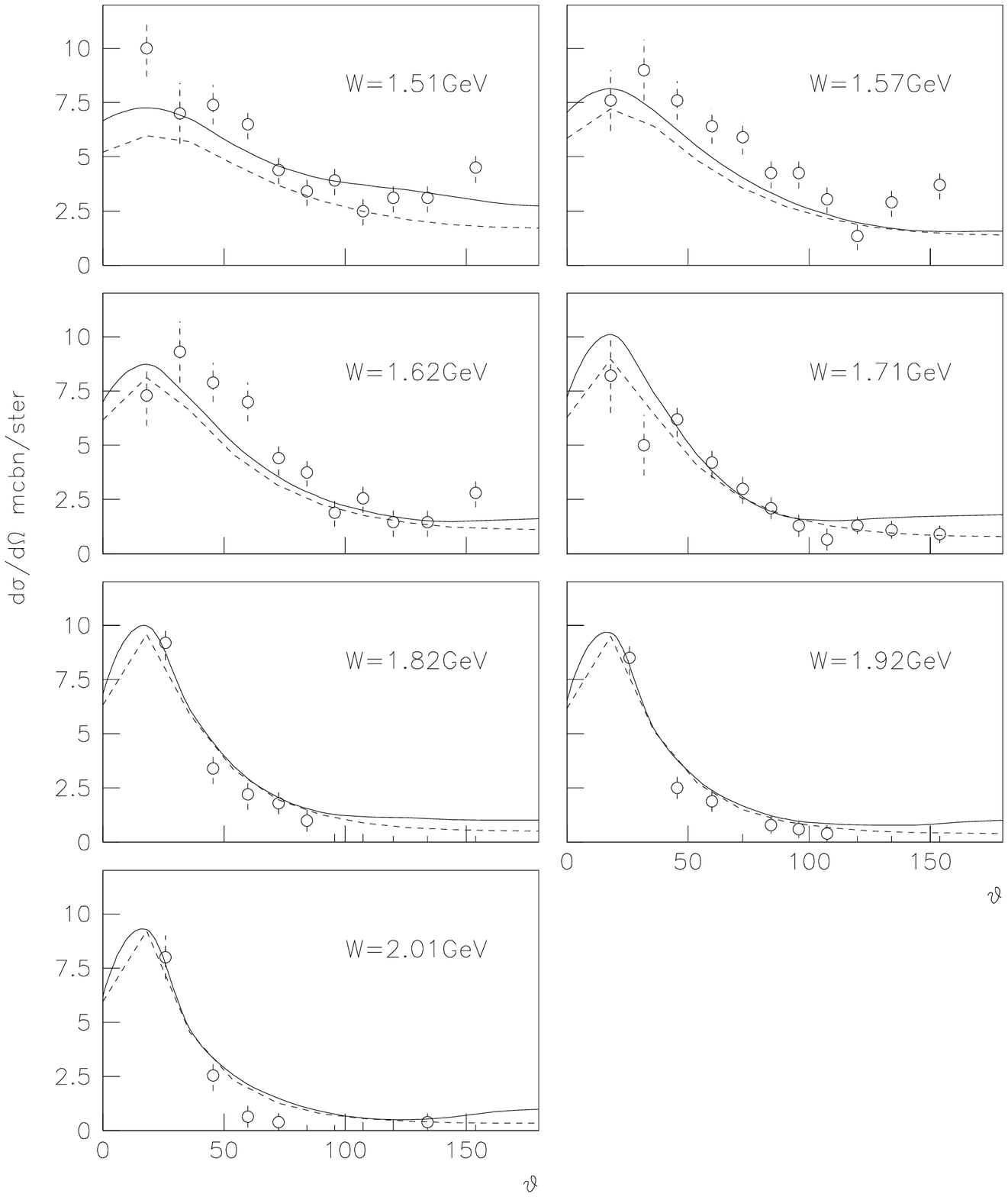,width=15cm}
Figure 6
\end{figure}

\newpage
\begin{figure}
\epsfig{file=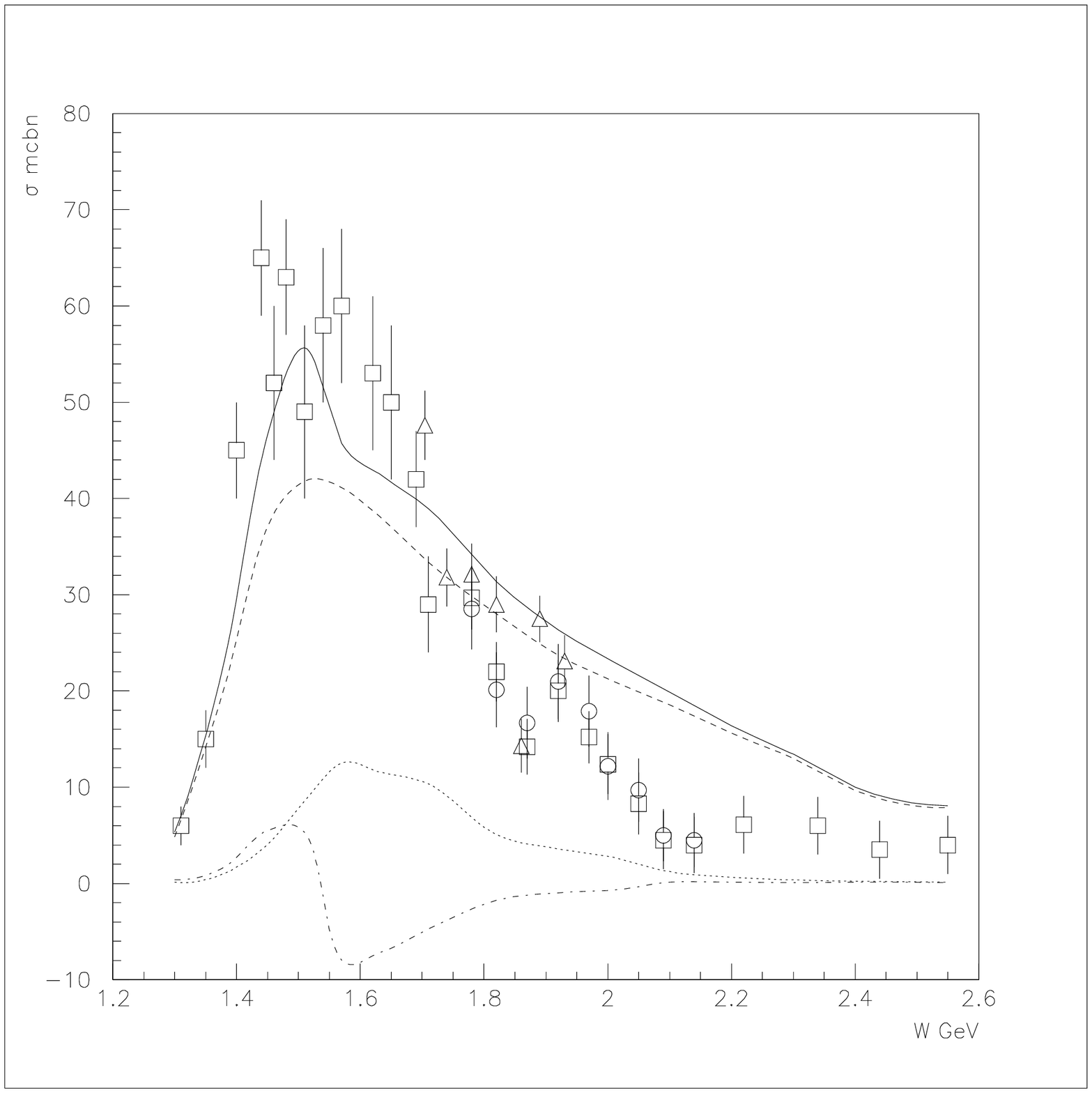,width=15cm}
Figure 7
\end{figure}

\newpage
\begin{figure}
\epsfig{file=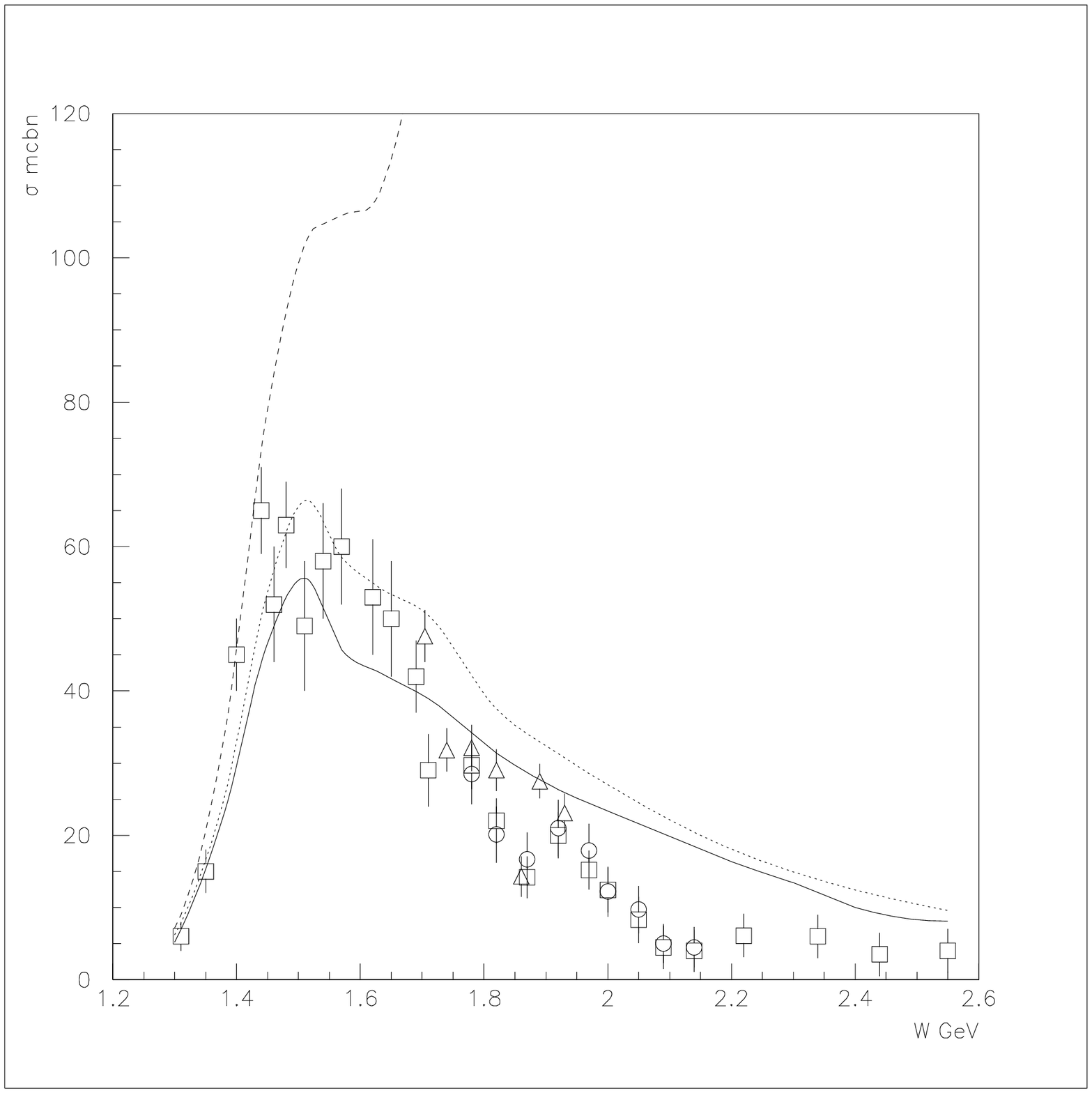,width=15cm}
Figure 8
\end{figure}

\newpage
\begin{figure}
\epsfig{file=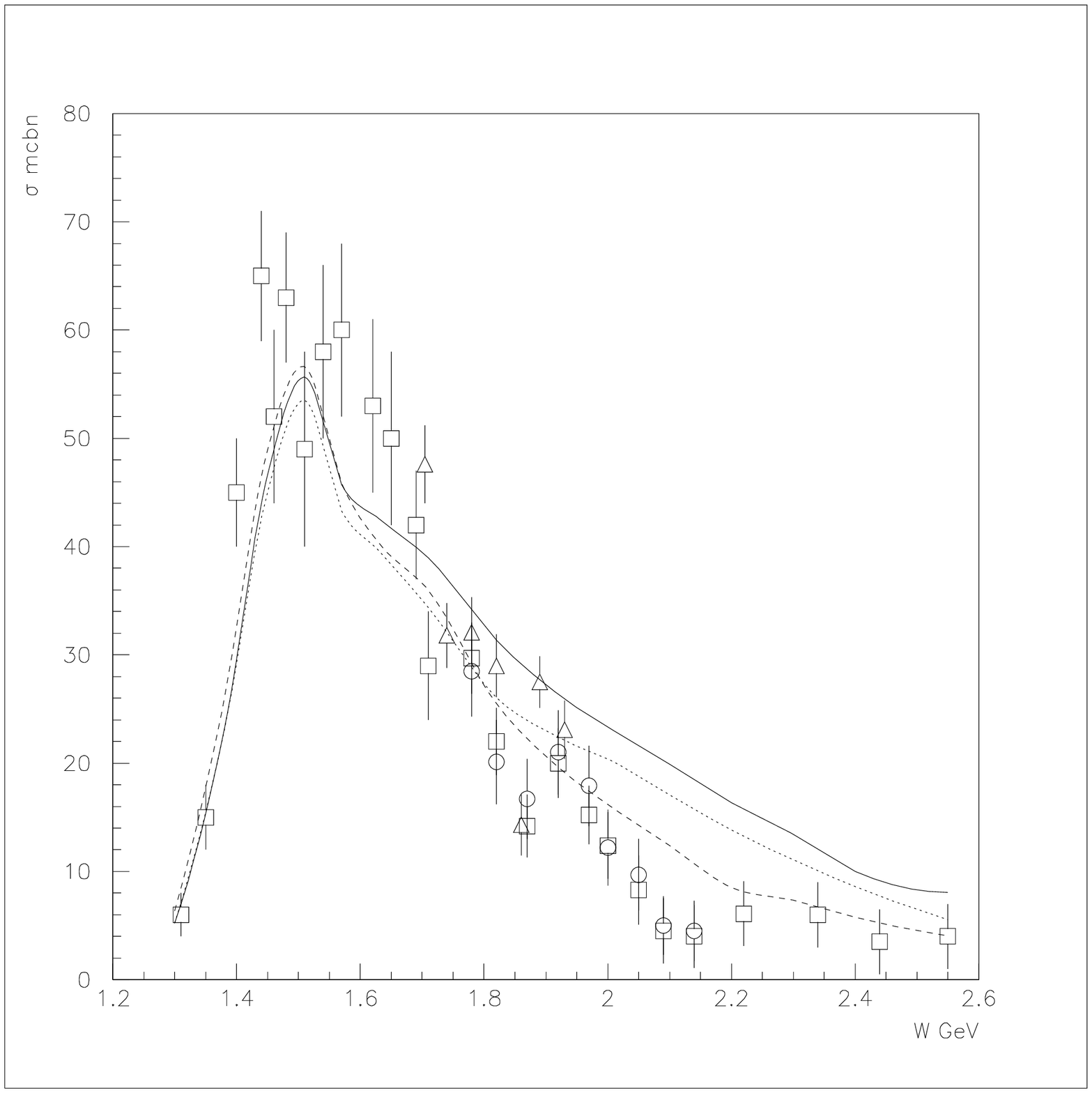,width=15cm}
Figure 9
\end{figure}

\newpage
\begin{figure}
\epsfig{file=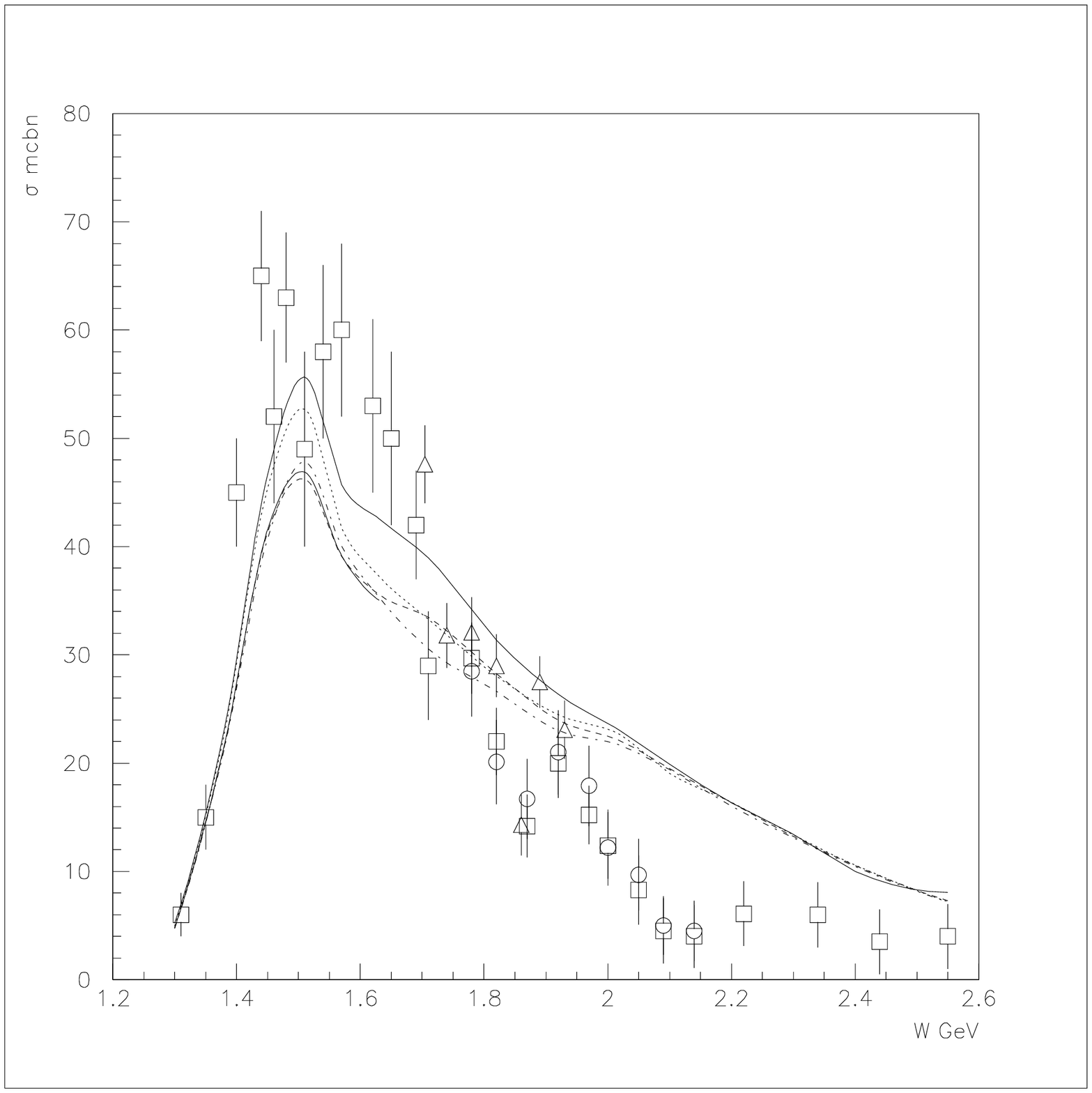,width=15cm}
Figure 10
\end{figure}

\begin{figure}
\epsfig{file=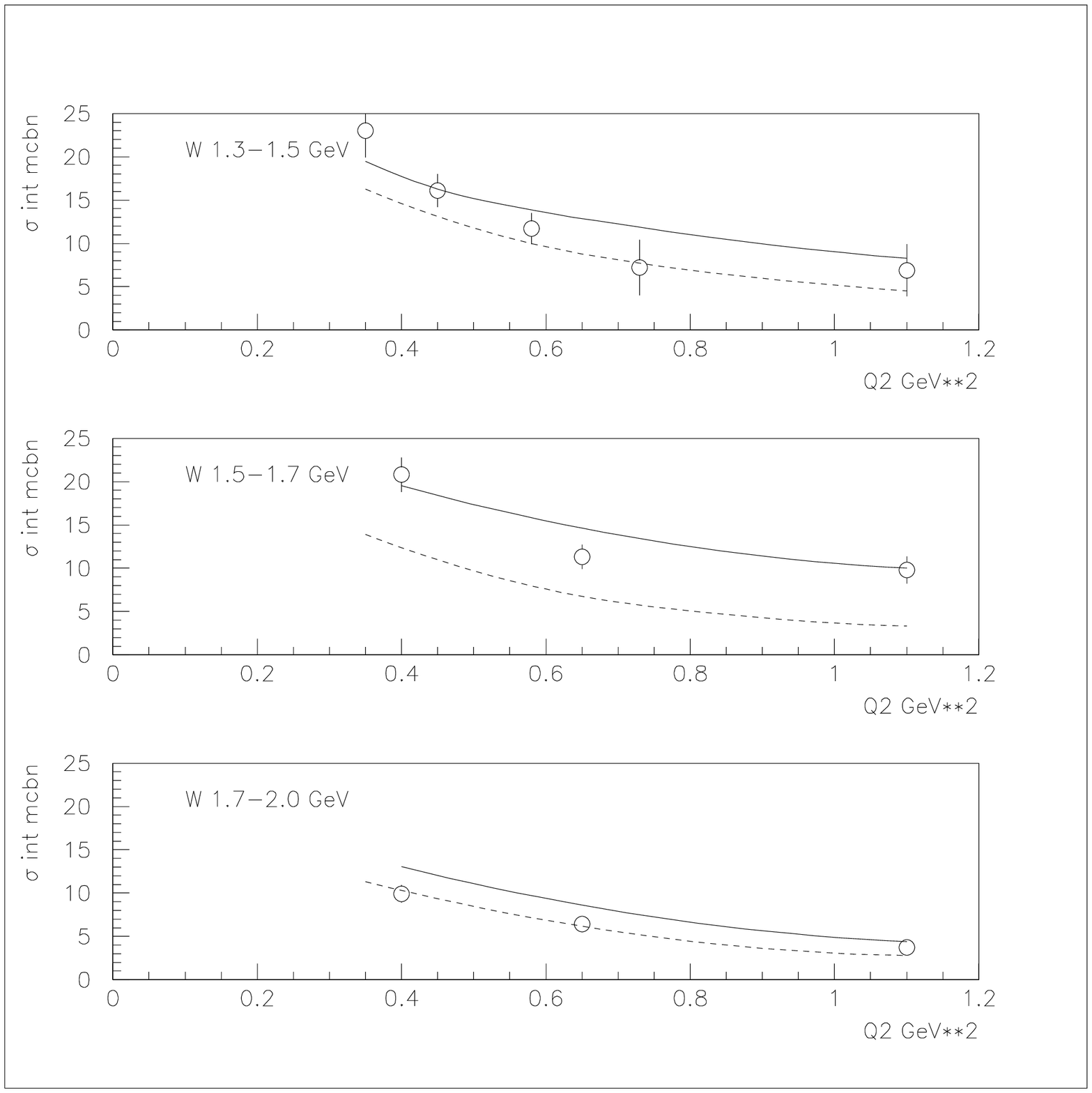,width=15cm}
Figure 11
\end{figure}

\newpage
\begin{figure}
\epsfig{file=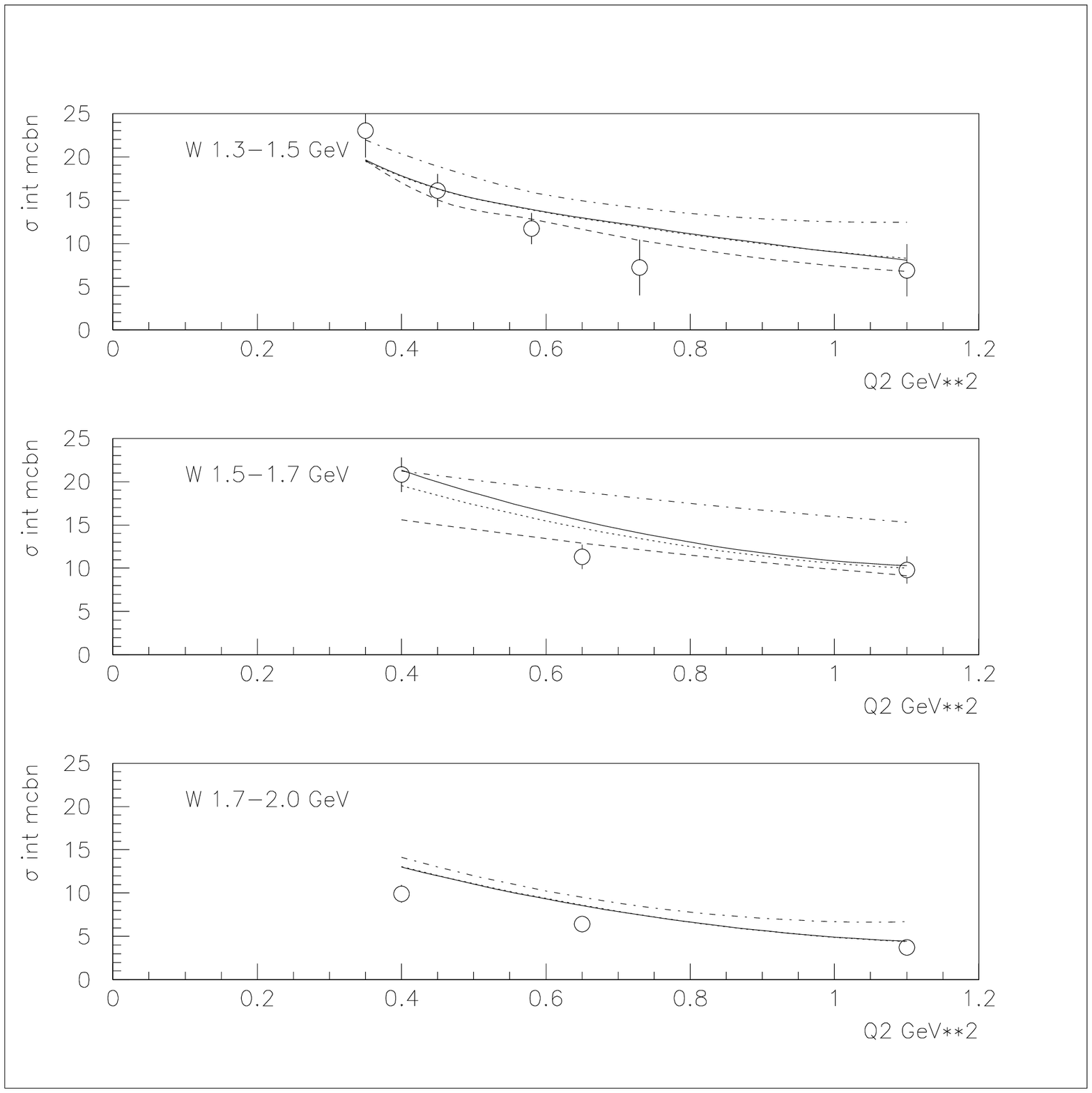,width=15cm}
Figure 12
\end{figure}

\end{document}